%% file: ismm20bCameraReady.tex
\documentclass[sigplan,screen]{acmart}

\AtBeginDocument{%
  \providecommand\BibTeX{{%
    \normalfont B\kern-0.5em{\scshape i\kern-0.25em b}\kern-0.8em\TeX}}}

\acmConference[ISMM '20]{Proceedings of the 2020 ACM SIGPLAN International Symposium on Memory Management}{June 16, 2020}{London, UK}
\acmBooktitle{Proceedings of the 2020 ACM SIGPLAN International Symposium on Memory Management (ISMM '20), June 16, 2020, London, UK}

\usepackage[redeflists]{IEEEtrantools}
\usepackage{subcaption}
\usepackage{multirow}
\usepackage{multicol}
\usepackage{balance}
\usepackage{microtype}
\usepackage{tabu}
\usepackage{color}
\newcommand{\thickhline}{%
    \noalign {\ifnum 0=`}\fi \hrule height 1pt
    \futurelet \reserved@a \@xhline
}
\definecolor{myblack}{rgb}{0,0,0}

\newcommand{\tech}{\text{{MNEME}}}{}
\newcommand{\ineq}[1]{\footnotesize$#1$\normalsize}{}

\begin{document}
\bstctlcite{IEEEexample:BSTcontrol}
\title{Exploiting Inter- and Intra-Memory Asymmetries for Data Mapping in Hybrid Tiered-Memories}

\author{Shihao Song}
\email{shihao.song@drexel.edu}
\affiliation{%
  \institution{Drexel University}
  \streetaddress{3101 Market Street}
  \city{Philadelphia}
  \state{PA}
  \country{USA}
}

\author{Anup Das}
\email{anup.das@drexel.edu}
\affiliation{%
  \institution{Drexel University}
  \streetaddress{3101 Market Street}
  \city{Philadelphia}
  \state{PA}
  \postcode{19104}
  \country{USA}
}

\author{Nagarajan Kandasamy}
\email{nk78@drexel.edu}
\affiliation{%
  \institution{Drexel University}
  \streetaddress{3101 Market Street}
  \city{Philadelphia}
  \state{PA}
  \postcode{19104}
  \country{USA}
}








\renewcommand{\shortauthors}{Shihao Song, Anup Das, and Nagarajan Kandasamy}

\begin{abstract}
\input{sections/abstract}
\end{abstract}




\maketitle

\section{Introduction}
\label{sec:introduction}
\input{sections/introduction.tex}

\section{New Segmented Bitline Architecture of \tech{}}
\label{sec:segmented_bitline}
\input{sections/segmented_bitline}


\section{New Page Management Policy of \tech{}}
\label{sec:mechanism}
\input{sections/mechanism.tex}

\section{Implementation of \tech{}}
\label{sec:implementation}
\input{sections/implementation.tex}

\section{Evaluation Methodology}
\label{sec:evaluation}
\input{sections/evaluation.tex}

\section{Results and Discussions}
\label{sec:results}
\input{sections/results.tex}

\section{Related Works}
\label{sec:related_works}
\input{sections/related_works}

\section{Conclusions}
\label{sec:conclusions}
\input{sections/conclusions.tex}


\balance
\bibliographystyle{IEEEtranSN}
\bibliography{mneme}


\end{document}

%% file: sections/abstract.tex
Modern computing systems are embracing hybrid memory comprising of DRAM and non-volatile memory (NVM) to combine the best properties of both memory technologies, achieving low latency, high reliability, and high density.
A prominent characteristic of DRAM-NVM hybrid memory is that it has NVM access latency much higher than DRAM access latency. We call this \emph{inter-memory asymmetry}.
We observe that parasitic components on a long bitline are a major source of high latency in both DRAM and NVM, and a significant factor contributing to high-voltage operations in NVM, which impact their reliability. We propose an architectural change, where each long bitline in DRAM and NVM is split into two segments by an isolation transistor.
One segment can be accessed with lower latency and operating voltage than the other.
By introducing tiers, we enable non-uniform accesses \textit{within} each memory type (which we call \textit{intra-memory asymmetry}), leading to performance and reliability trade-offs in DRAM-NVM hybrid memory.

We show that our hybrid tiered-memory architecture has a tremendous potential to improve performance and reliability, if exploited by an efficient page management policy at the operating system (OS). Modern OSes are already aware of inter-memory asymmetry. They migrate pages between the two memory types during program execution, 
starting from an initial allocation of the page to a \textit{randomly-selected} free physical address in the memory. We extend existing OS awareness in \textit{three} ways. \textit{First}, we \textit{exploit} both inter- and intra-memory asymmetries to allocate and migrate memory pages between the tiers in DRAM and NVM.
\textit{Second}, we 
improve the OS's page allocation decisions by \textit{predicting} the access intensity of a newly-referenced memory page in a program and placing it to a \textit{matching tier} during its initial allocation. This minimizes page migrations during program execution, lowering the performance overhead. 
\textit{Third}, we propose a solution to migrate pages between the tiers of the same memory without transferring data over the memory channel, minimizing channel occupancy and improving performance. Our overall approach, which we call \tech{}, to enable and exploit asymmetries in DRAM-NVM hybrid tiered memory improves both performance and reliability for both single-core and multi-programmed workloads. 

%% file: sections/introduction.tex
\begin{figure}[h!]
	\centering
	\centerline{\includegraphics[width=0.99\columnwidth]{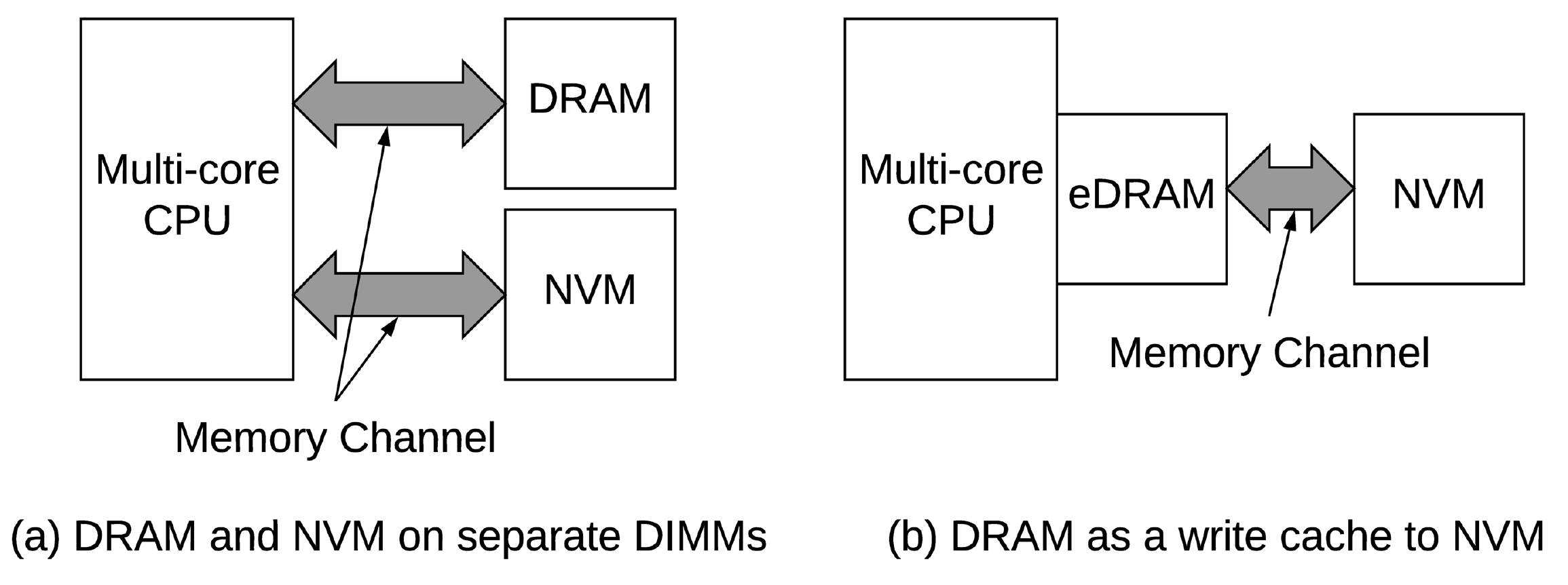}}
	\vspace{-10pt}
	\caption{DRAM-NVM hybrid memory architecture of (a) 3D XPoint Memory~\cite{bourzac2017has} and (b) IBM POWER9~\cite{sadasivam2017ibm}.}
	\vspace{-10pt}
	\label{fig:hybrid_memory}
\end{figure}

DRAM has long been the choice substrate for architecting main memory subsystems due to its low cost per bit. However, DRAM is a fundamental performance and energy bottleneck in almost all computer
systems~\cite{wulf1995hitting,wilkes2001memory,mutlu2015research,mutlu2013memory}, and is experiencing significant technology scaling challenges~\cite{mutlu2013memory,kang2014co,mutlu2017rowhammer,mutlu2019rowhammer}. DRAM-compatible emerging non-volatile memory (NVM) technologies such as Flash \cite{cai2017error}, oxide-based RAM (OxRAM) \cite{mallik2017design}, phase-change memory (PCM) \cite{wong2010phase}, and spin transfer torque magnetic RAM (STT-MRAM) \cite{apalkov2013spin} can address some of these challenges~\cite{LeeISCA09,kultursay2013evaluating,song2019enabling,songISMMa,ren2015thynvm,lee2010phase2,QureshiISCA09}. However, they are usually slower than DRAM and have limited endurance.\footnote{NVM's endurance ranges from $10^5$ writes for Flash to $10^{10}$ writes for OxRAM, and PCM in between, with $10^7$ writes.}
Modern computing systems are therefore embracing hybrid memory designs comprising of DRAM and NVM \cite{nation2009computer}. These systems
combine the best properties of both memory technologies to improve latency, reliability, capacity, and cost. The non-volatile 3D XPoint memory~\cite{bourzac2017has} is one example of a hybrid memory, with DRAM and NVM connected to separate channels, interfacing with a multi-core CPU chip using the JEDEC's new NVDIMM specification~\cite{jedecnvdimm2017}.
IBM POWER9 architecture~\cite{sadasivam2017ibm} is another example, which uses embedded DRAM (eDRAM) as a write cache to NVM-based main memory. Figure~\ref{fig:hybrid_memory} illustrates both these hybrid architectures and we evaluate them in Section~\ref{sec:results}.

Modern operating systems (OSes) such as Nimble~\cite{yan2019nimble} are already aware of the performance and reliability asymmetry in hybrid memory. They migrate write-intensive pages to DRAM (which has practically \textit{infinite} endurance) and read-intensive pages to NVM (which has a read latency \textit{comparable} to DRAM), starting from an initial \textit{random} page placement~\cite{blagodurov2019hot}. There are two key limitations in these OSes. First, if pages are not placed in their matching memory (i.e., NVM or DRAM) at their initial allocation, they 
can incur significant performance and energy overhead during program execution due to the high bank and channel occupancy in moving page data between the two memory. Second, limited write endurance is not the only reliability issue in NVM. In fact, a recent study has shown that even read accesses can lead to high-voltage related aging (another key reliability issue) in a NVM's peripheral circuit~\cite{balajical19}.

Our \textbf{objective} is to improve performance and reliability (both endurance and aging) of DRAM-NVM hybrid memory. We achieve this goal by exploiting the following \textit{three} major observations in this paper.

\textbf{Observation 1:} 
\textit{
A significant number of pages are migrated more than once during a program execution. 
}

\begin{figure}[h!]
	\centering
	\centerline{\includegraphics[width=0.99\columnwidth]{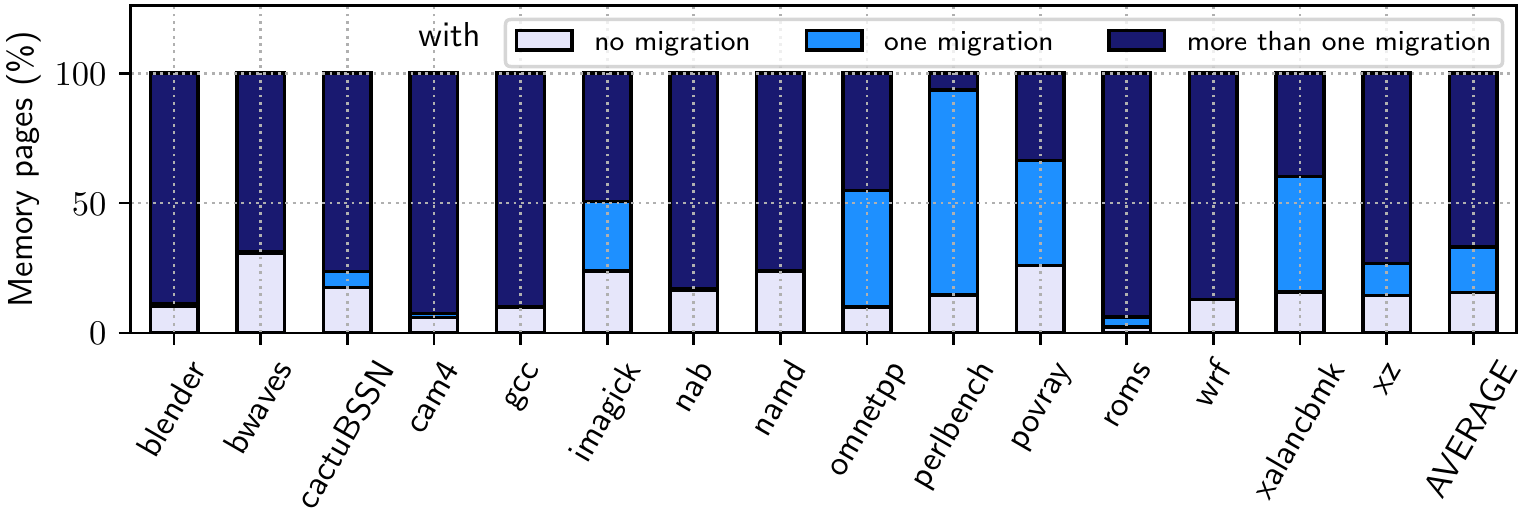}}
	\vspace{-10pt}
	\caption{Fraction of memory pages that suffer no migration, exactly one migration, and more than one migration using Nimble for our evaluated workloads.}
	\vspace{-10pt}
	\label{fig:migration_all}
\end{figure}

Figure~\ref{fig:migration_all} plots the fraction of memory pages with no migration, exactly one migration, and more than one migration using Nimble's dynamic page migration policy for the evaluated workloads, which are detailed in Section~\ref{sec:evaluation}.

We observe a wide variation in behavior across these programs. For instance, over 93\% of memory pages in perlbench are migrated at most once, whereas 95\% of all pages in roms are migrated more than once. On average, 67\% of memory pages in these programs suffer more than one migration during their execution. These migrations lead to high energy and performance overhead.

Observation 1 leads to our \textbf{first key idea} that if a memory page is placed in a matching memory during its initial allocation, many of these migrations can be eliminated, leading to performance and energy improvements (Section \ref{sec:results}). This idea also leads to our next observation.

\textbf{Observation 2:} 
\textit{
There are typically only a few first-touch instructions (FTIs) in a program and only a small percentage of these instructions induce the most memory accesses. 
}

Modern OSes implement first-touch page allocation policy, where a virtual-to-physical address translator allocates a random physical memory page from the free pool to a virtual page address, when the virtual page is first touched by a memory instruction in the program. We call this memory instruction \textit{first-touch instruction} (\textbf{FTI}).

\begin{figure}[h!]
	\centering
	\centerline{\includegraphics[width=1.01\columnwidth]{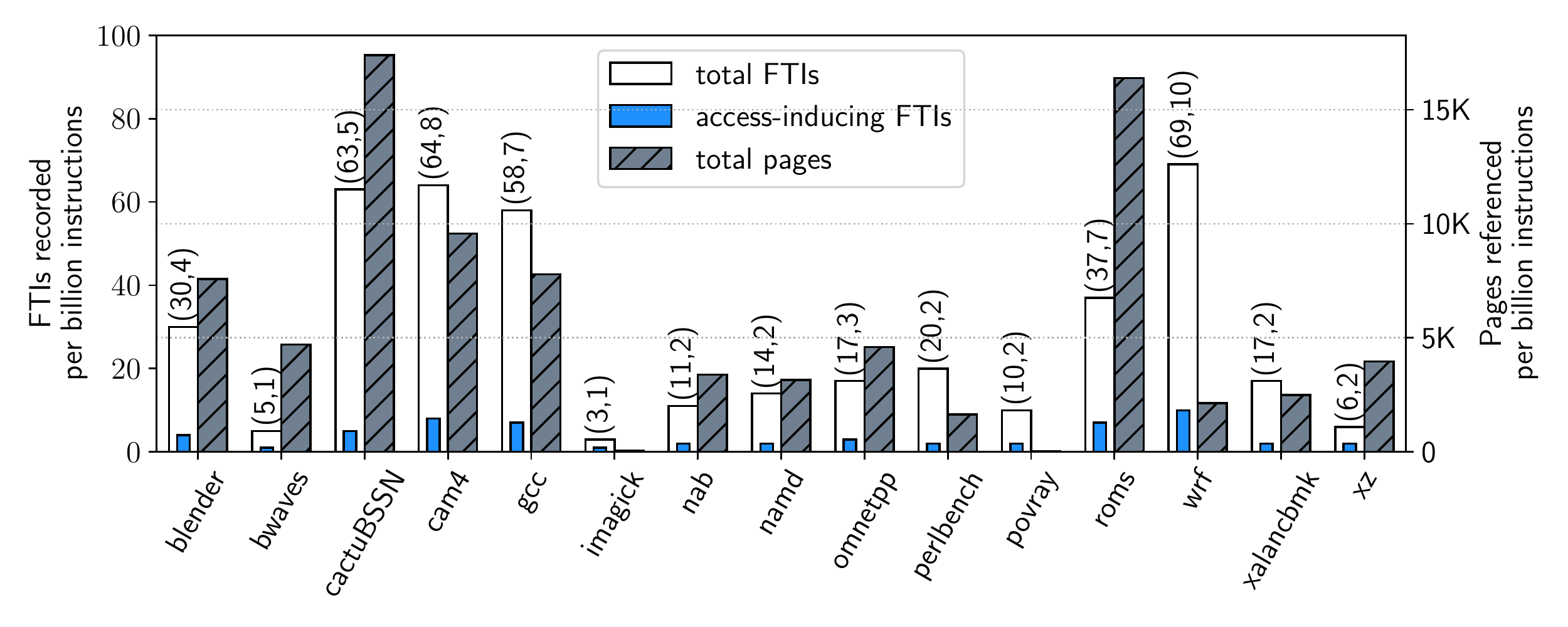}}
	\vspace{-10pt}
	\caption{First-touch instructions (FTIs) per billion instructions of the evaluated workloads.}
	\vspace{-5pt}
	\label{fig:ftis}
\end{figure}

Figure \ref{fig:ftis} plots the number of FTIs and referenced pages per billion instructions of the evaluated workloads. We report total FTIs (outer first bar in each set) and the number of FTIs that touch pages which serve over 90\% of memory accesses (inner first bar). We also report the pages referenced per billion instructions (second bar). We observe that 1) there are very few FTIs per billion instructions of each workload, and 2) on average, only 17\% of FTIs in a program touch pages which serve over 90\% of memory accesses.

Observation 2 leads to our \textbf{second key idea} of profiling FTIs based on the number of accesses to memory pages they touch and using it to predict the access intensity of a newly-referenced memory page, thereby placing it in a matching memory during its initial allocation.

Observations 1 and 2 are related to OS-based page management in DRAM-NVM hybrid memory. Our final observation is related to the \textit{internal architecture} of memory.

\textbf{Observation 3:} 
\textit{
Parasitic components on a long bitline are a major source of high latency in both DRAM and NVM, and a significant factor contributing to high-voltage operations in NVM, which impact reliability. 
}

This is observed and reported for DRAM \cite{lee2013tiered}. We expand this observation for NVM,  where each bit is represented by the resistance of a cell (low resistance represents logic `1' and high resistance logic `0'). 
An NVM cell's resistance is read or programmed by driving current through the cell using a peripheral circuit, which consists of sense amplifiers (to read) and write drivers (to write). We analyze the internal architecture of an NVM bank and find that its peripheral circuit is several orders of magnitude larger than the size of an NVM cell~\cite{song2019enabling,cho2005programming,dray2018high,sandhu2018memory}.\footnote{NVM, like DRAM, is organized hierarchically. An example NVM of 128GB capacity can have 4 {channels}, with 4 ranks per channel, and 8 banks per rank. A bank can have {8} {partitions}, which are similar to subarrays in DRAM~\cite{KimISCA12}.} To amortize this large size, memory designers connect a peripheral circuit to many (typically 4096) NVM cells through a wire called a \textit{bitline}. Numerous bitlines are laid in parallel to form an NVM array (called a \emph{tile}). A row of NVM cells is called a \emph{wordline}. 

\begin{figure}[h!]
	\centering
	\centerline{\includegraphics[width=0.99\columnwidth]{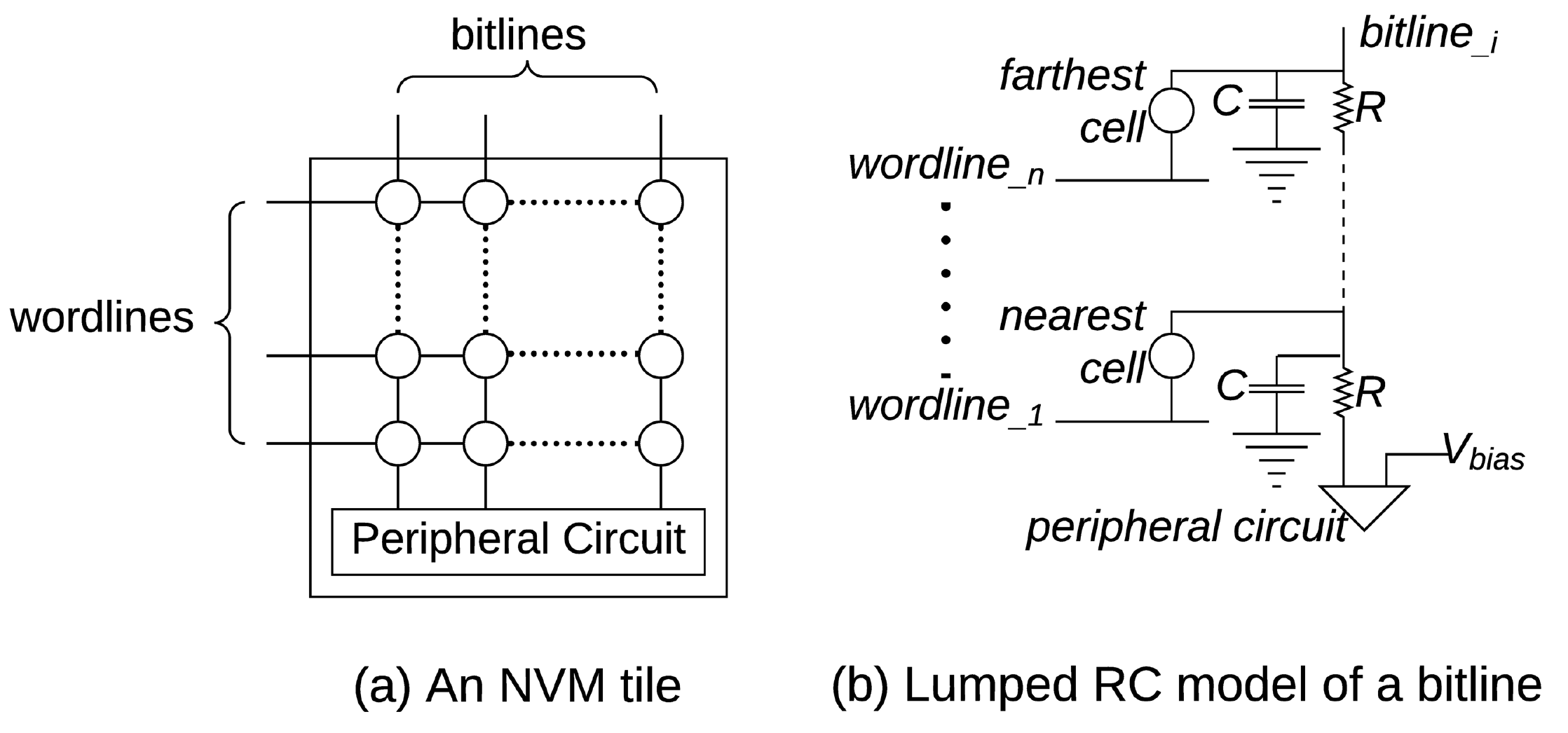}}
	\vspace{-10pt}
	\caption{(a) An NVM tile with bitlines and wordlines and (b) Lumped RC model of a bitline.}
	\vspace{-10pt}
	\label{fig:tile_cell_structure}
\end{figure}

Figure~\ref{fig:tile_cell_structure} (a) illustrates an NVM tile with bitlines and wordlines. Many such tiles make a partition (see our simulation parameters in Table~\ref{tab:simulation_parameters}). Figure~\ref{fig:tile_cell_structure}(b) illustrates the lumped RC circuit of a bitline to model its parasitic components. The voltage drop (called the \textit{IR drop}) on the bitline parasitic needs to be compensated by a peripheral circuit to access the NVM cells on its bitline. As we can see from this figure, farther a cell from the peripheral circuit, higher is the IR drop.

\begin{figure}[h!]
	\centering
	\centerline{\includegraphics[width=0.99\columnwidth]{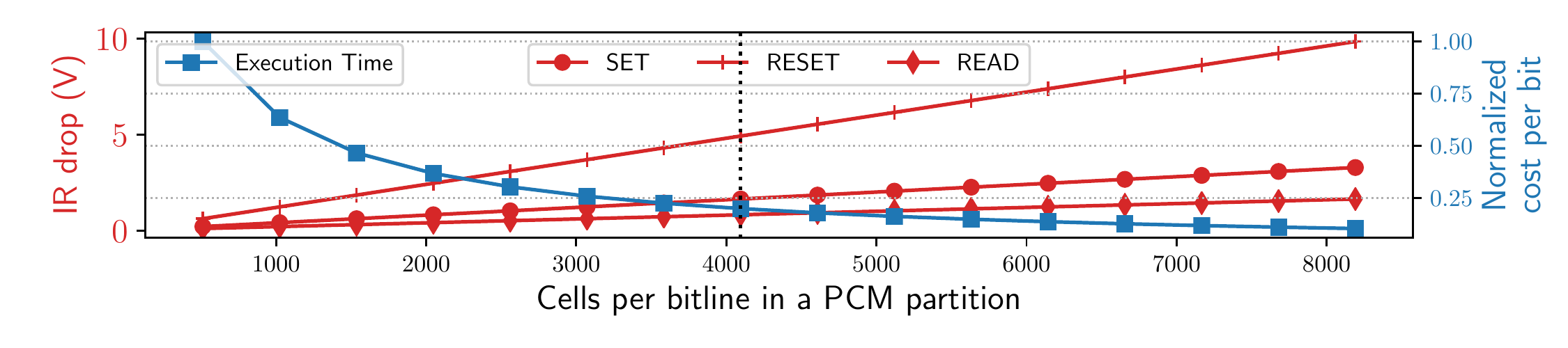}}
	\vspace{-10pt}
	\caption{Architecting the number of PCM cells per bitline.}
	\vspace{-5pt}
	\label{fig:cost_per_bit}
\end{figure}

Figure~\ref{fig:cost_per_bit} plots the design analysis performed while architecting main memory. We show such analysis for PCM, an emerging NVM, based on Micron's 45nm design~\cite{bhattacharyya2019memory}.\footnote{We expect the values to be of similar orders of magnitude for other designs.} The left y-axis plots the IR drop on a bitline for SET, RESET, and READ operations as a function of the number of bitline cells. The right y-axis plots the normalized cost per bit. We observe that the normalized cost decreases as the number of bitline cells increases. However, higher the number of bitline cells, higher is the IR drop.
The trade-off point is typically set to 4096 cells in most PCM designs~\cite{bhattacharyya2019memory,lung2016double,redaelli2017semiconductor,villa2018pcm}. Similar analysis conducted on Micron's 45nm DRAM design suggests that 512 cells per bitline in a DRAM subarray gives the best latency and cost trade-offs~\cite{lee2013tiered,das2018vrl}. 
In this paper, we assume, without loss of generality, each bitline in NVM and DRAM contains 4096 and 512 cells, respectively. 

Table~\ref{tab:bias} reports the biasing voltage needed to access the nearest cell (1${}^{\text{st}}$ cell), the farthest cell (4096${}^{\text{th}}$ cell) and an intermediate cell (512${}^{\text{th}}$) on a bitline in PCM. Higher voltages are needed to access cells that are farther from their peripheral circuit. This has two implications. First, cells that are nearer to their peripheral circuit can be accessed faster. This is because the on-chip voltage regulator, which supplies the biasing voltage for a peripheral circuit, has faster response time and higher energy efficiency to generate lower voltages. Second, operating a peripheral circuit at a lower voltage incurs lower circuit aging, which improves reliability (see our reliability formulation in Section \ref{sec:nvm_reliability}).
We conclude that PCM (and in general, NVM) has asymmetric latency and reliability in accessing its content.

\begin{table}[h!]
    \caption{PCM's biasing voltages.}
	\label{tab:bias}
	\vspace{-5pt}
	\centering
	\setlength{\tabcolsep}{2pt}
	{\fontsize{8}{10}\selectfont
    \begin{tabular}{c|ccc}
    \hline
    \multirow{3}{*}{\textbf{Cell Op.}} & \multicolumn{3}{|c}{\textbf{Bias Voltage}} \\ \cline{2-4}
    & \textbf{Nearest cell} & \textbf{Farthest cell} & \textbf{Intermediate cell}\\
    & \textbf{(1${}^\text{st}$ cell)} & \textbf{(4096${}^\text{th}$ cell)} & \textbf{(512${}^\text{th}$ cell)}\\
    \hline
    \textit{SET} & 2.1 & 3.7V & 2.3V  \\
    \textit{RESET} & 6.8 & 7.1V & 6.9V\\
    \textit{READ} & 0.96 & 2.85V & 1.2V \\
    \hline
    \end{tabular}}
\end{table}


Observation 3 leads to our \textbf{third key idea} of introducing an isolation transistor on each bitline in DRAM and NVM, to allow its length to appear shorter when accessing cells nearer to its peripheral circuit, thereby achieving low latency in DRAM and NVM and additionally, high reliability in NVM. Segmented bitlines create latency and reliability asymmetry, i.e., tiers within both DRAM and NVM.

We introduce \textbf{\tech{}}\footnote{In Greek mythology, \tech{} is the muse of memory. \tech{} means persistent effect of memory of past events, which are the first-touch instructions in the context of this paper.}, 
a mechanism that
builds on the three ideas above to \textit{enable} 
additional tiers in hybrid memory and \textit{exploit} these tiers through an efficient OS-level page allocation policy.
\tech{} places a newly-referenced memory page to the best tier during its initial allocation, minimizing channel and bank occupancy associated with migration of page data during execution. 
Through \tech{}, we make the following key \textbf{contributions}.
\begin{itemize}
    \item We introduce a new memory architecture with segmented bitlines within DRAM and NVM to improve performance and reliability during data accesses.
    \item We propose an approach to predict access intensity of a newly-referenced memory page using the page's first-touch instruction (FTI) and 
    place it in a matching memory tier during its initial allocation, reducing page migrations during program execution.
    
    \item We develop our FTI-based page allocation for the entire program duration to adapt to and make correct allocation decisions for different phases of execution in a program with potentially distinct working sets.
    
    \item We show how to reduce channel occupancy during page migration between tiers of the same memory, thereby improving performance.
    
    \item We introduce an efficient hardware implementation of \tech{} using Bloom filters.
    
    \item We implement \tech{} for two hybrid memory architectures and also for commodity DRAM-based mainstream architecture, and show significant performance and reliability improvements for both single-program and multi-programmed workloads.
\end{itemize}

%% file: sections/segmented_bitline.tex
\begin{figure}[h!]
	\centering
	\centerline{\includegraphics[width=0.99\columnwidth]{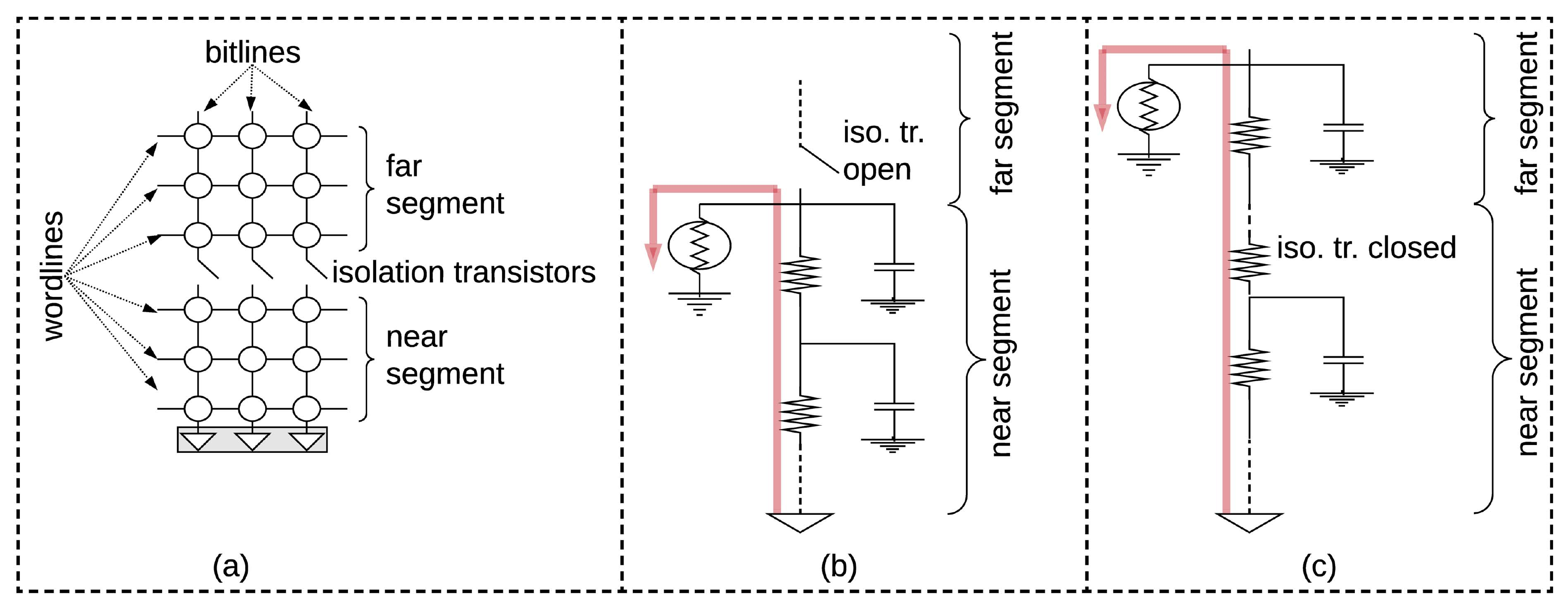}}
	\vspace{-5pt}
	\caption{(a) Bitlines partitioned into segments. (b) Accessing a near segment cell. (c) Accessing a far segment cell.}
	\vspace{-5pt}
	\label{fig:splitted_bitline}
\end{figure}

Figure~\ref{fig:splitted_bitline}(a) shows the proposed segmented bitline architecture, where each long bitline in DRAM and NVM is split into two segments using an isolation transistor: the segment connected directly to the peripheral circuit is called the \textit{near segment}, whereas the other is called the \textit{far segment}. 
Cells in the near segment can be accessed faster using lower bias voltages due to the reduced parasitic on the current path (see Figure~\ref{fig:splitted_bitline}(b)). This improves performance. Additionally, by using lower voltages, circuit aging when accessing the near segment is minimized, which improves reliability. Aging-related reliability is particularly critical for NVM, which requires higher operating voltages than DRAM \cite{songISMMa}.

However, the isolation transistor increases access latency of the far segment and introduces 
its ON resistance 
in the current path, which imposes additional bias requirement for the peripheral circuit (see Figure~\ref{fig:splitted_bitline}(c)). This lowers reliability of the peripheral circuit in accessing the far segment. 

Segmented bitlines is previously proposed for DRAM subarrays~\cite{lee2013tiered} and they lead to performance trade-offs in accessing near versus far segments. We propose segmented bitlines for DRAM-NVM hybrid memory, which introduces reliability trade-off, in addition to the performance ones.

\begin{figure}[h!]
	\centering
	\vspace{-5pt}
	\centerline{\includegraphics[width=0.99\columnwidth]{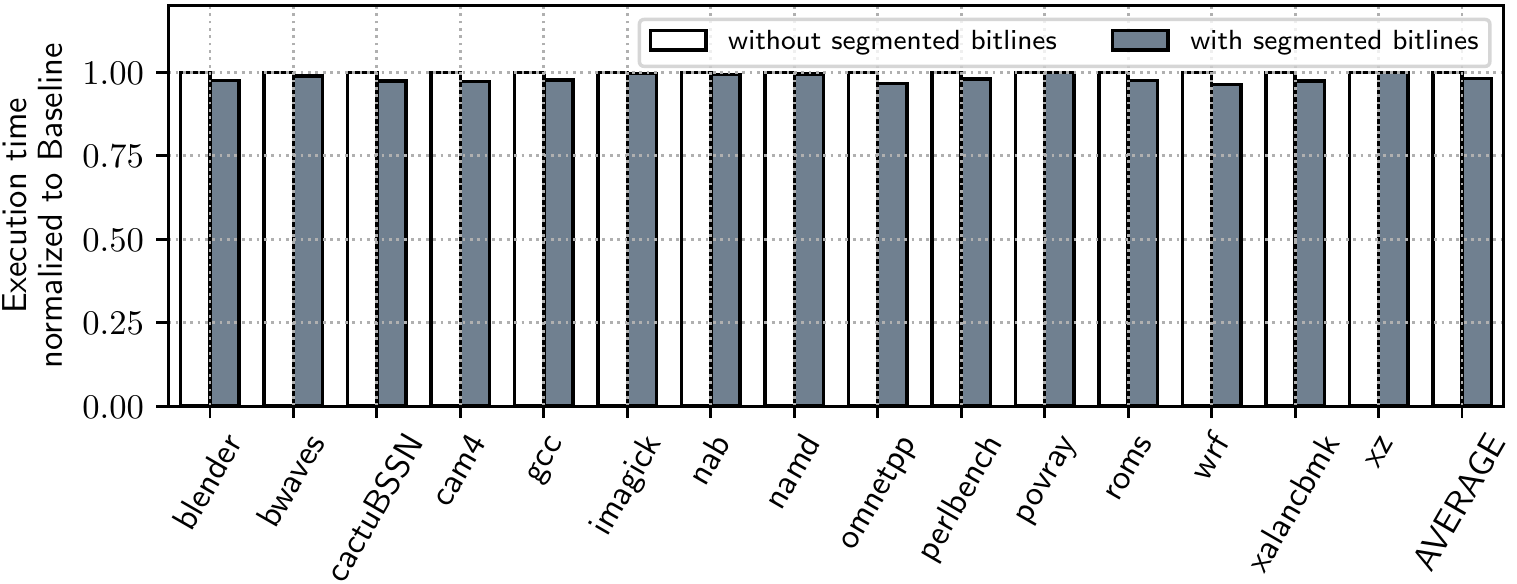}}
	\vspace{-10pt}
	\caption{Execution time of our evaluated workloads normalized to Baseline where bitlines are not segmented.}
	\vspace{-5pt}
	\label{fig:observation_1}
\end{figure}

To evaluate the performance improvement using this new memory architecture, Figure~\ref{fig:observation_1} plots the execution time of 15 workloads (see Section~\ref{sec:evaluation}) on a hybrid memory with segmented bitlines. Results are normalized to the execution time of a Baseline design, where bitlines are not segmented. We observe that simply introducing memory tiers by creating segments in each bitline is just not enough to guarantee performance improvement; in fact, performance improves by only 2\% on average for these workloads.
We believe that our hybrid tiered-memory architecture can only deliver on its promises if the inter- and intra-memory asymmetries are exploited efficiently by 
an operating system (OS)-level page management policy, which we introduce next.

\begin{figure}[h!]
	\centering
	\centerline{\includegraphics[width=0.89\columnwidth]{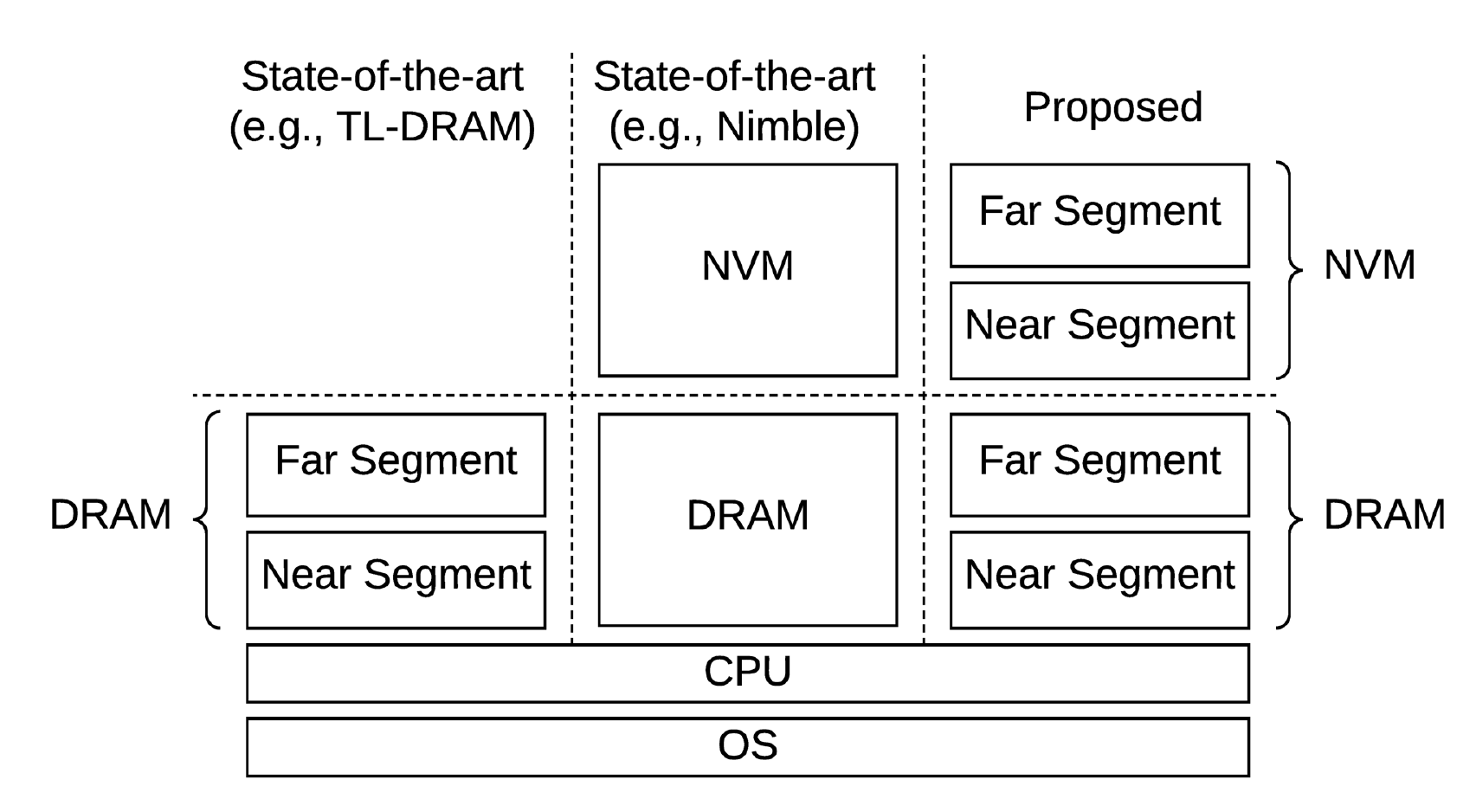}}
	\vspace{-5pt}
	\caption{Proposed hybrid tiered-memory architecture.}
	\label{fig:tms}
\end{figure}

Our {hybrid tiered-memory architecture} is shown in Figure~\ref{fig:tms}, where the memory tiers are arranged with increasing access latency from the CPU. The figure also shows how our architecture differs from the two state-of-the-art approaches: TL-DRAM~\cite{lee2013tiered}, which only uses memory tiers within DRAM-based main memory and Nimble~\cite{yan2019nimble}, which uses DRAM-NVM hybrid main memory like ours but the bitlines are \textit{not} segmented. We evaluate both these state-of-the-art approaches in Section \ref{sec:results}.

%% file: sections/mechanism.tex
\begin{figure}[h!]
	\centering
	\centerline{\includegraphics[width=0.69\columnwidth]{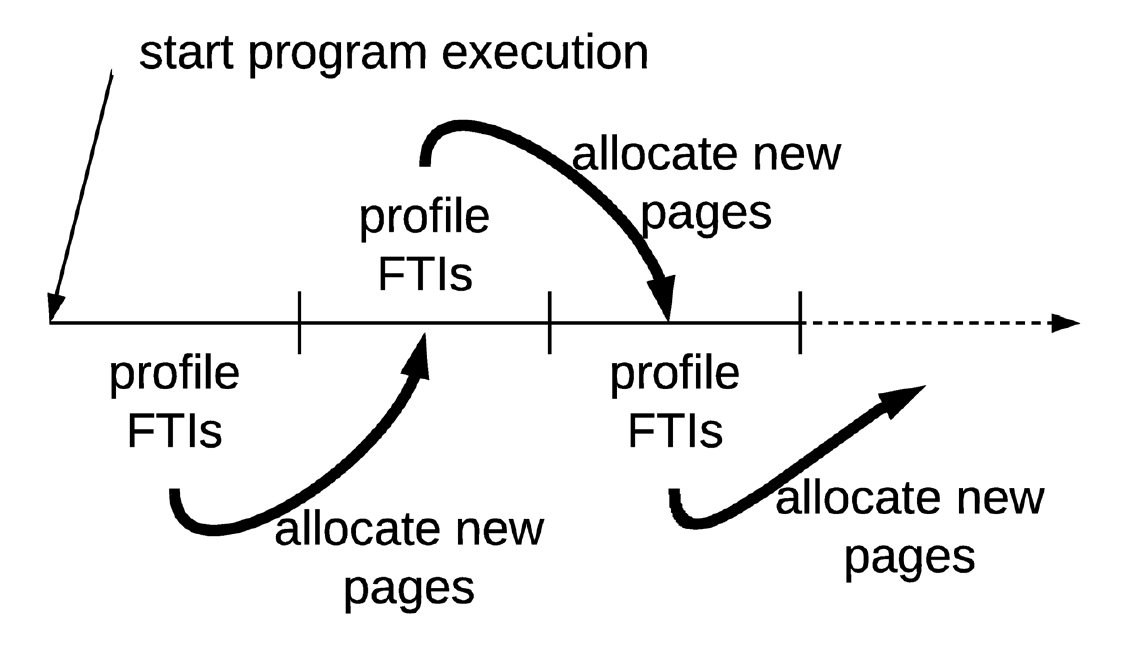}}
	\vspace{-10pt}
	\caption{Proposed program execution.}
	\vspace{-10pt}
	\label{fig:program_execution}
\end{figure}

Figure~\ref{fig:program_execution} shows a high-level overview of our page allocation policy to exploit our tiered architecture. A program is broken down into fixed intervals (called \emph{phases}). At each phase,
we profile FTIs based on accesses to pages they touch.
This information is then used to decide the initial placement of all newly-referenced memory pages of the subsequent phases.

\input{sections/nk_notes.tex}

\begin{figure}[h!]
	\centering
	\centerline{\includegraphics[width=0.99\columnwidth]{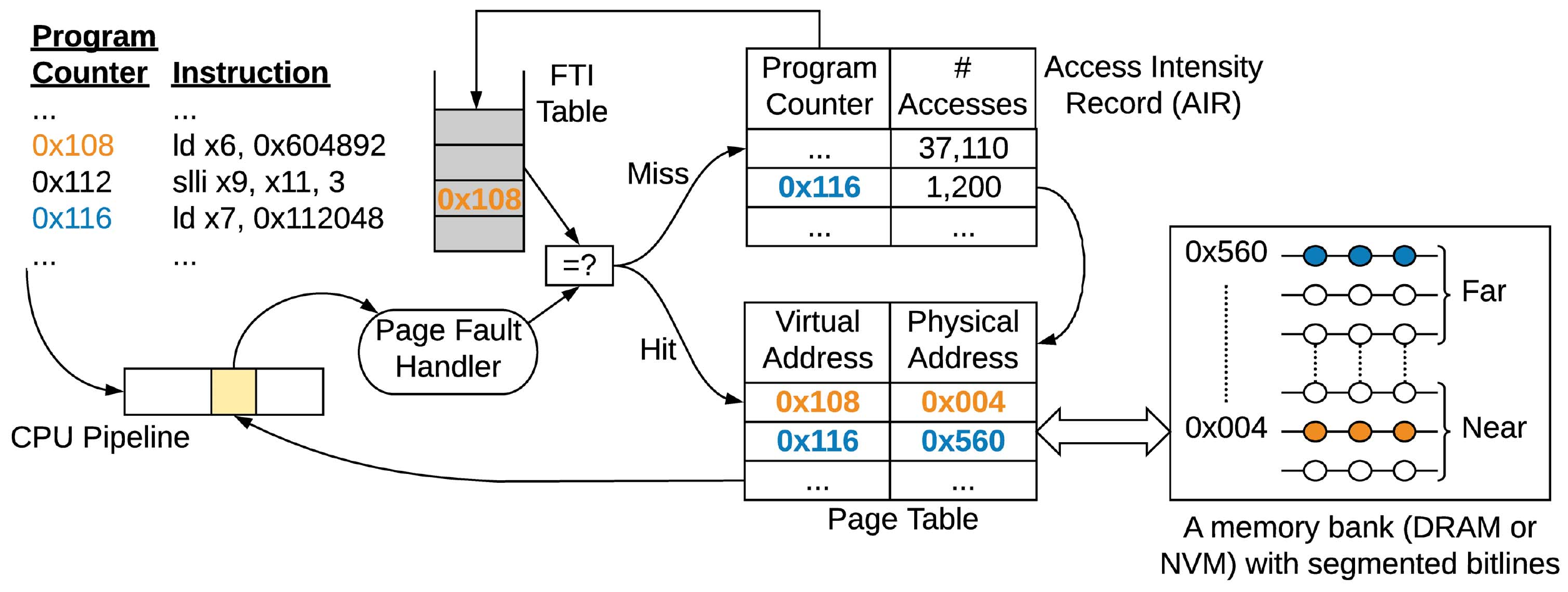}}
	\vspace{-10pt}
	\caption{A conceptual overview of \tech{}.}
	\vspace{-10pt}
	\label{fig:overview}
\end{figure}

A \textbf{conceptual overview} of \tech{} is shown in Figure~\ref{fig:overview}. At a high-level, the memory controller maintains a table containing the memory addresses of access-intensive first-touch instructions (i.e., their program counter value). 
We call this \textbf{FTI table}. We split the execution of an application into phases, with each phase comprising of 100 million instructions (See Section~\ref{sec:interval} for evaluation on the size of an execution phase). 

To allocate a {newly} referenced memory page in an execution phase, the OS page-fault handler runs a custom instruction to check if the virtual address corresponding to the FTI of the page hits in the FTI table. \textit{If a match is found}, 
the memory page is predicted to be access-intensive. The page-fault handler allocates this new memory page to the near memory segment. We explain later how to choose between DRAM and NVM. The memory controller then uses reduced timing and voltage parameters to access this page, improving performance and reliability. \textit{Otherwise}, the FTI is considered to be unknown and possibly referencing a non-access intensive memory page. The OS page handler allocates the memory page to the far segment, while tracking the number of accesses this page generates within the phase, leveraging OS page tracking structures, and recording it inside a table. We call this \textbf{Access Intensity Record (AIR)}. At the end of each phase, the top access-intensive unknown FTIs of AIR (with number of accesses higher than a threshold) are inserted into the FTI table to predict and place all new memory pages to near or far segments. In this way, the FTI table is constantly updated with new access-inducing FTIs that are uncovered during program execution.


Using the FTI table and AIR, \tech{} can place a new memory page to a specific segment in DRAM or NVM. To select the specific memory type (i.e., DRAM vs. NVM), we introduce the following changes: 1) we maintain two FTI tables: one holding those FTIs that touch more write-intensive pages (we call this \emph{FTI\_W}), and another holding those FTIs that touch more read-intensive pages (we call this \emph{FTI\_R}), and
2) extend the AIR to record the number of read-inducing and write-inducing pages that each FTI touches. 

For a new memory page in a program phase, there can be four possibilities with the corresponding FTI.
\begin{itemize}
    \item \emph{Hit in FTI\_W and miss in FTI\_R:} the FTI is predicted as write-access inducing. So, allocate the memory page to a near segment in DRAM.
    \item \emph{Miss in FTI\_W and hit in FTI\_R:} the FTI is predicted as read-access inducing. So, allocate the memory page to a near segment in NVM.
    \item \emph{Hit in FTI\_W and hit in FTI\_R:} the FTI is predicted as both read and write inducing. So, allocate the memory page to a near segment in DRAM (conservative).
    \item \emph{Miss in FTI\_W and miss in FTI\_R:} the FTI is predicted as non-access inducing. So, allocate the memory page this FTI touches to a far segment in NVM if space is available there, otherwise allocate it to a far segment in DRAM. Additionally, make an entry for the FTI in AIR and start recording accesses to the page.
\end{itemize}


If \tech{} predicts the access intensity of a new memory page correctly (which we evaluate in Section~\ref{sec:results}), the page will be placed in the correct memory tier during its initial allocation, reducing run-time page migration overhead.
Otherwise, the page will be placed in an incorrect tier and will be migrated by tracking its accesses during program execution. 



\begin{figure}[h!]
 	\centering
 	\centerline{\includegraphics[width=0.99\columnwidth]{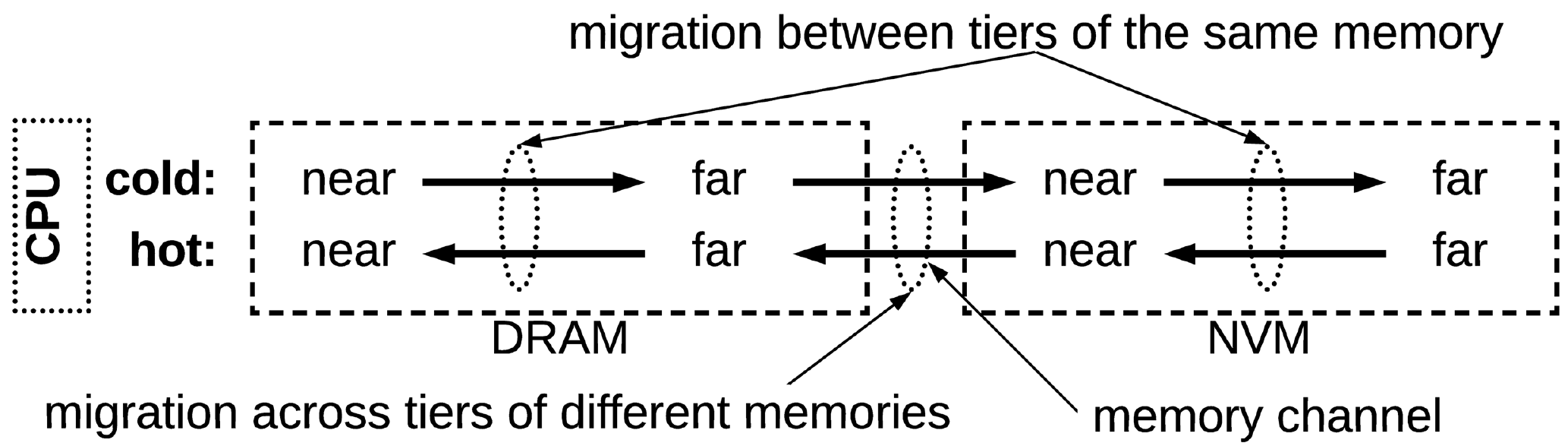}}
 	\vspace{-10pt}
 	\caption{Data migration in \tech{}.}
 	\label{fig:migration_overview}
\end{figure}

\textbf{Page migration in \tech{}:} Figure \ref{fig:migration_overview} shows the journey of hot and cold pages through tiers of the proposed hybrid tiered-memory architecture. \tech{} supports two types of migrations: 1) page migrations between tiers of the same memory unit, and 2) page migrations across tiers of different memory units. For within memory migrations, we propose an approach where a page can be migrated from one tier to another in the same memory bank without utilizing the memory channels. This can be achieved for DRAM using two back-to-back activates (see Sec. \ref{sec:lat}) utilizing row buffers, which are \textit{shared} between read and write operations (see RowClone \cite{seshadri2013rowclone} for instance). 

\begin{figure}[h!]
 	\centering
 	\centerline{\includegraphics[width=0.89\columnwidth]{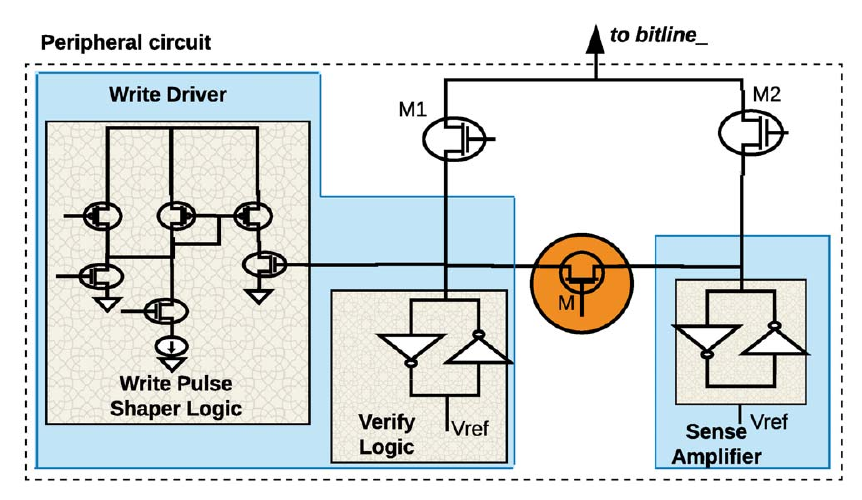}}
 	\vspace{-10pt}
 	\caption{{Internal circuit of a bank's peripheral structure.}}
 	\label{fig:peripheral_circuit_overview}
\end{figure}

However, for PCM, and NVM in general, this is not straight forward because the peripheral circuit consists of \textit{separate hardware} to read and write. 
Figure \ref{fig:peripheral_circuit_overview} shows the architecture of a peripheral circuit in an NVM (e.g., PCM) bank \cite{song2019enabling}. The peripheral circuit consists of the sense amplifier (to read) and the write driver (to write), which are connected to a bitline using transistors \texttt{M1} and \texttt{M2}. 
From the write driver's internal circuit diagram shown in Figure \ref{fig:peripheral_circuit_overview}, we observe that the write driver can be viewed as a collection of two components -- the \textbf{write pulse shaper logic}, which generates the current pulses necessary for the cell's SET and RESET operations, and the \textbf{verify logic}, which verifies the correctness of these operations. These two circuit components together serve write requests from the bank using a write scheme known as \emph{program-and-verify}  (P\&V)~\cite{nirschl2007write,goda2018programming}. 

Based on this observation, we propose \emph{simple} circuit modifications to introduce the decoupling transistor \texttt{M} (see Fig. \ref{fig:peripheral_circuit_overview}), which can be configured when needed, to transfer data from the sense amplifier to the verify logic. As a result of this modification, we can program the data read in the sense amplifier to a different row in the bank using the write driver. This facilitates data migration between tiers of the same memory bank without utilizing external memory channels.

To migrate pages across tiers of different memory units, we still use the memory channel, which leads to performance overhead. However, our OS-level page allocation policy minimizes these migrations significantly (see Section \ref{sec:overheads}).



%% file: sections/nk_notes.tex
\begin{figure}[h!]
	\centering
	\centerline{\includegraphics[width=0.7\columnwidth]{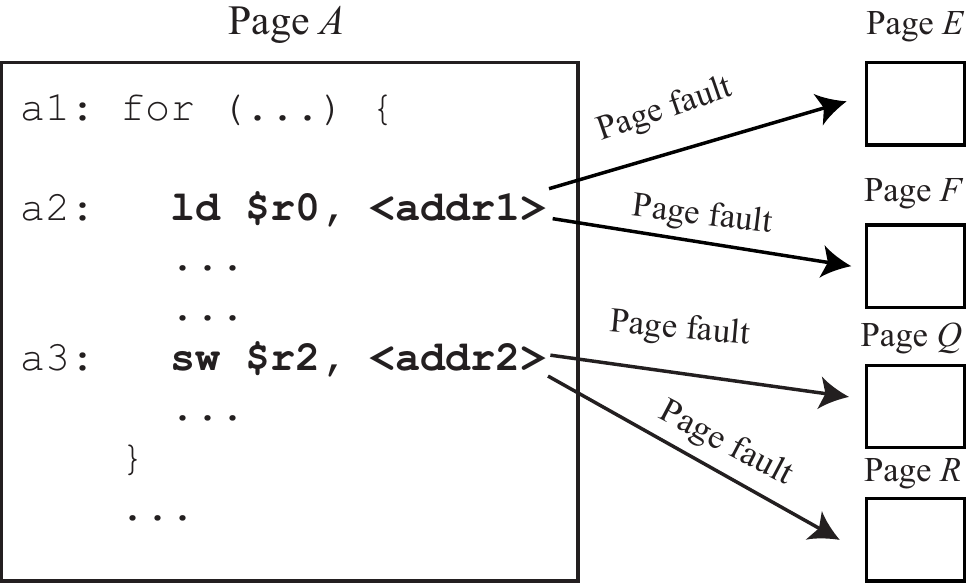}}
	\vspace{-10pt}
	\caption{Examples of subsequent memory accesses created by first-touch load and store instructions.}
	\vspace{-10pt}
	\label{fig:fti_examples}
\end{figure}

Figure~\ref{fig:fti_examples} shows an {example} of using FTIs to predict the access intensity of newly-referenced memory pages in \tech{}. 

Assume that no profile information is available in the beginning.
The program counter \texttt{a1} causes the initial page fault, loading Page $A$ into the far memory segment.
During the course of program execution, the load instruction at \texttt{a2} causes a page fault, loading page $E$ into the far memory segment. Similarly, the store instruction at \texttt{a3} also loads page $Q$ into the far segment. The instructions at addresses, \texttt{a1}, \texttt{a2}, and \texttt{a3} are the FTIs. Now assume that as we iterate through the for loop, the load and store instructions, located at addresses \texttt{a2} and \texttt{a3}, respectively, generate numerous accesses to $E$ and $Q$, respectively. \tech{} records them as FTIs that induce large numbers of accesses to any page that these instructions might load in the future. Therefore, later on when the load instruction references an address that requires page $F$ to be loaded, MNEME will load this page into the near memory segment. Similarly, $R$ will also be loaded in the near segment when accessed by the store instruction.   

In addition to load and store statements, implementations of branch or jump tables will also contain FTIs that can be predicted to load frequently-accessed pages; for example, when a jump instruction is the FTI that causes the page containing the corresponding function to be loaded.
Fig.~\ref{fig:ftis} shows that such FTIs are a major source of memory references.

%% file: sections/implementation.tex
\tech{} consists of two key components: 1) interface to support efficient data tiering within and across memory units in hybrid memory, and 2) intensity prediction via FTI table and AIR. We discuss how to implement each of these components in order to design an efficient implementation of \tech{}.

\subsection{Interface to Support Tiered Hybrid Memory}
\label{sec:architecture}
\input{sections/interface.tex}

%% file: sections/interface.tex
\begin{figure}[h!]
 	\centering
 	\centerline{\includegraphics[width=0.99\columnwidth]{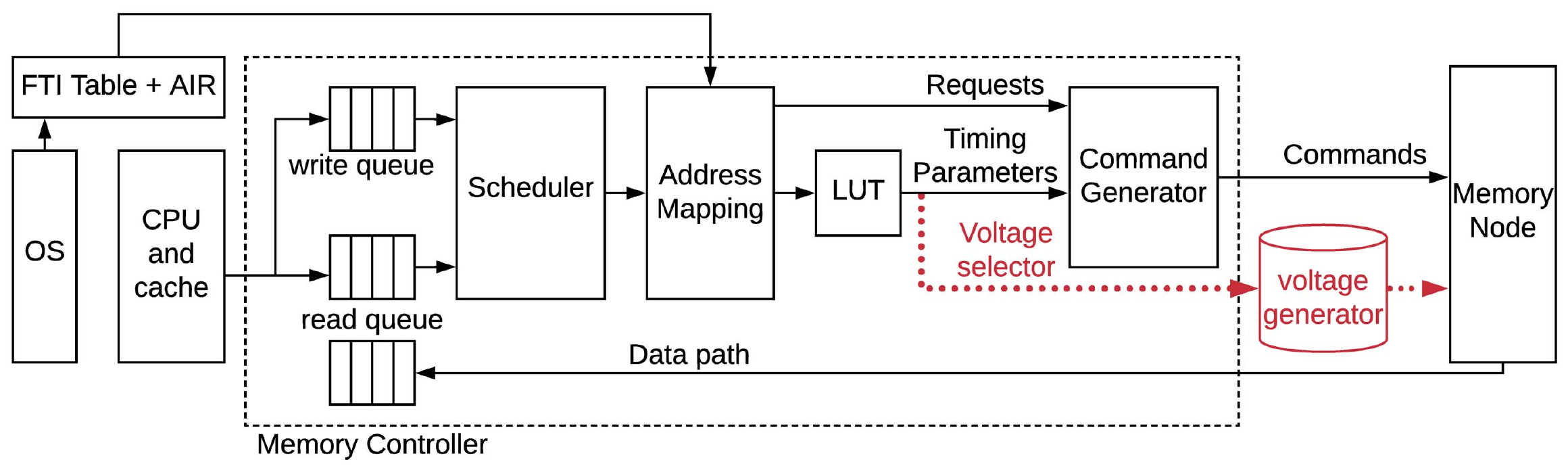}}
 	\vspace{-10pt}
 	\caption{A full-system overview.}
 	\label{fig:system_overview}
\end{figure}

Figure~\ref{fig:system_overview} shows the handshaking between OS, CPU and memory, via the memory controller.
Inside the memory controller, there is a separate read and write queue to buffer requests to the memory. The scheduler schedules requests from these queues using its access scheduling policy. We use the FR-FCFS policy~\cite{RixnerISCA2000}, where the scheduler prioritizes requests that hit in the row buffer in a memory bank. 

The address mapping block is responsible for selecting the desired timing and voltage parameters based on the memory segment (near or far) that a request hits. To explain this, we use an example of 128GB NVM with 4 channels, 4 ranks per channel, 8 banks per rank, 8 partitions per bank, 128 tiles per partition, 4096 wordlines and 2048 bitlines per tile. Considering bank interleaving, the address mapping scheme is as follows: [36:35] = rank address, [34:32] = partition address, [31:25] = tile address, [24:13] = row address, [12:11] = column address, [10:8] = bank address, [7:6] = channel address, and [5:0] = byte address. We assume, without loss of generality, that an isolation transistor divides each bitline into near segment with 512 cells and far segment with 3584 cells (= 4096 - 512).

To decode the segment that a request hits, we use bit slicing on the wordline address bits [24:13]. Therefore, address bit 22 is used as the segment select bit (`0'\ineq{\implies} near segment and `1'\ineq{\implies} far segment). The segment select bit is used to select segment-specific timing parameters stored in a lookup table (LUT) inside the memory controller. The memory request and the selected timing parameters are forwarded to a command generator, implemented as a state machine, which issues memory-specific commands at appropriate intervals.

The description above applies to DRAM as well, with the exception of the on-chip voltage regulator, which is needed only for NVM to drive current through its cells. Furthermore, we use the DRAM configuration of Lee et al.~\cite{lee2013tiered}, where a bitline contains 512 cells and is divided using an isolation transistor into near segment with 128 cells and far segment with 384 cells. Bit slicing is performed accordingly.

We now provide latency and reliability analysis of near and far segments in DRAM and NVM. We consider the DRAM architecture of Lee et al.~\cite{lee2013tiered} and the PCM architecture of Redaelli~\cite{pcm_book}, both from Micron.

\subsubsection{Latency Analysis}
\label{sec:lat}
To understand the latency impact due to bitline segmentation, we briefly review memory-timing parameters. The following discussion applies for both DRAM and PCM. To serve a memory request that accesses data at a particular row and column address within a bank, a memory controller issues \emph{three} commands to the bank.
\begin{itemize}
	\item \texttt{\underline{ACTIVATE}:} activate the wordline and enable the peripheral circuit for the memory cells to be accessed.
	\item \texttt{\underline{READ}/\underline{WRITE}:} drive read or write current through the cell (PCM) or share charge from the cell (DRAM). After this command executes, the data stored in the cell is available at the output terminal of peripheral circuit, or the write data is programmed to the cell.
	\item \texttt{\underline{PRECHARGE}:} deactivate the wordline and bitline, and prepare the bank for the next access. 
\end{itemize}

\begin{figure}[h!]
 	\centering
 	\centerline{\includegraphics[width=0.9\columnwidth]{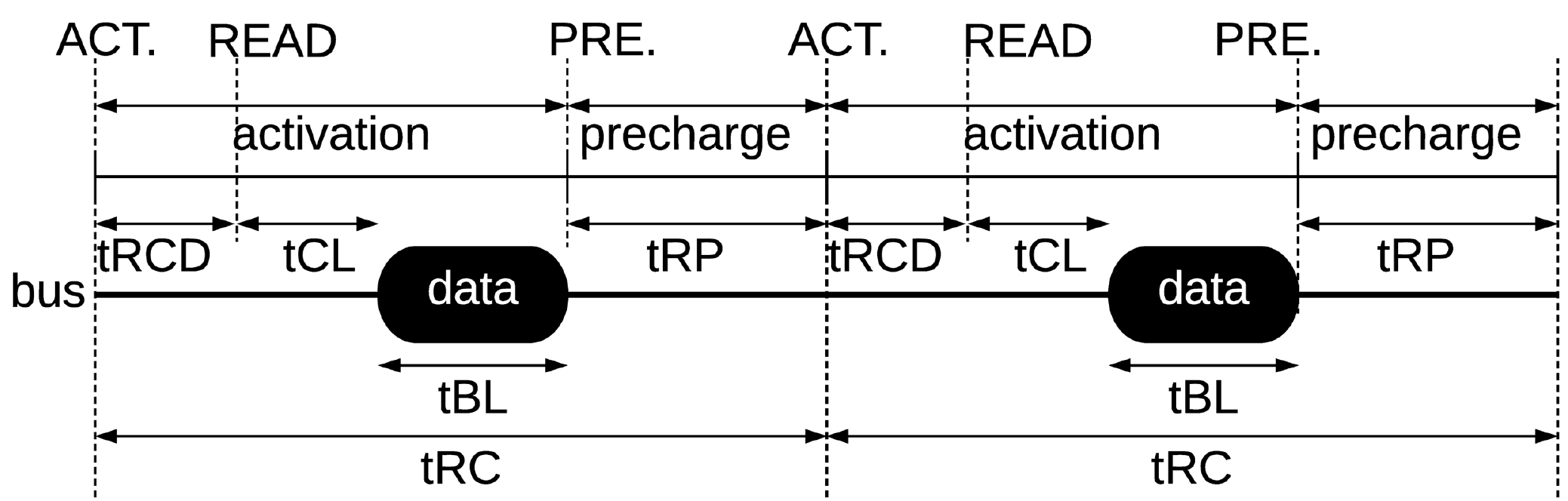}}
 	\vspace{-10pt}
 	\caption{Memory timings for read requests.}
    \vspace{-10pt}
 	\label{fig:memory_timings}
\end{figure}


Figure~\ref{fig:memory_timings} shows different memory timing parameters when serving two read requests. Table~\ref{tab:min_max_latency} reports these timing parameters for the near and far segment of DRAM and PCM. The parameters for DRAM are obtained from Lee et al.~\cite{lee2013tiered}, scaled to 45nm technology nodes using predictive technology scaling~\cite{cao2007mosfet}. The timing parameters for PCM are obtained via SPICE simulations~\cite{antognetti1990semiconductor} with 45nm PDK~\cite{stine2007freepdk}. 
Table \ref{tab:min_max_latency} is stored in a LUT in our memory controller.


\begin{table}[h!]
\setlength{\tabcolsep}{5pt}
\renewcommand{\arraystretch}{1.0}
\caption{Latency incurred by read and write requests, respectively, to near and far segments of DRAM and PCM.}
\vspace{-5pt}
\label{tab:min_max_latency}
\centering
{\fontsize{8}{10}\selectfont
\begin{tabu}{c c c c c c c c}
    \tabucline[2pt]{-}
    \multicolumn{3}{c}{} & \textbf{tRCD} & \textbf{tCL} & \textbf{tBL} & \textbf{tRP} & \textbf{tRC}\\
    \hline
    \multirow{4}{*}{DRAM} & \multirow{2}{*}{near} & Read & 9.3ns & 5.5ns & 7.5ns & 5.5ns & 27.8ns \\\cline{3-8}
     &  & Write & 9.3ns & 5.5ns & 7.5ns & 5.5ns & 27.8ns \\\cline{2-8}
     & \multirow{2}{*}{far} & Read & 15ns & 15ns & 7.5ns & 15ns & 52.5ns \\\cline{3-8}
     & & Write & 15ns & 15ns & 7.5ns & 15ns & 52.5ns \\
     \hline
    \multirow{4}{*}{PCM} & \multirow{2}{*}{near} & Read & 3.75ns & 22.5ns & 15ns & 0ns & 41.25ns \\\cline{3-8}
     &  & Write & 3.75ns & 101ns & 15ns & 0ns & 119.75ns \\\cline{2-8}
     & \multirow{2}{*}{far} & Read & 3.75ns & 37.5ns & 15ns & 0ns & 56.25ns \\\cline{3-8}
     & & Write & 3.75ns & 142.8ns & 15ns & 0ns & 161.55ns \\
    \tabucline[2pt]{-}
\end{tabu}
}
\end{table}


\subsubsection{Reliability Analysis}
\label{sec:nvm_reliability}
Table~\ref{tab:reliability_summary} summarizes the sources of reliability concerns in NVM.
In this work, we consider two dominant reliability issues in PCM: 1) finite endurance of PCM cells and 2) high voltage-related aging of CMOS devices in a peripheral circuit. We formulate these next.

\begin{table}[h]
\renewcommand{\arraystretch}{1.5}
\setlength{\tabcolsep}{2pt}
\caption{Reliability issues in NVM.}
\label{-10pt}
\label{tab:reliability_summary}
\centering
{\fontsize{8}{10}\selectfont
\begin{tabular}{|l|c|}
\hline
\textbf{Reliability Issues} & \textbf{NVM}\\
\hline
High-voltage related circuit aging & PCM, Flash\\
High-current related circuit aging & OxRAM, STT-MRAM\\
Read disturbance & All\\
Limited endurance & All\\
\hline
\end{tabular}}
\end{table}


\noindent\emph{\underline{Endurance-related lifetime}:} Endurance-related lifetime dep\-ends on: 1) how many times a PCM cell can be programmed (\ineq{N_e}) and 2) how frequently the cells are programmed (\ineq{N_f})~\cite{QureshiISCA12}. If \ineq{N_{WL}} is the total number of wordlines in a PCM bank, the endurance-related lifetime can be estimated as
\begin{equation}
    \label{eq:endurance_lifetime}
    \footnotesize {L_e} = N_{WL} * N_e / N_f.
\end{equation}

\noindent\emph{\underline{Aging-related lifetime}:} High-voltage operations lead to reliability issues such as negative-bias temperature instability (NBTI), hot carrier injection (HCI), and time-dependent dielectric breakdown (TDDB) \cite{das2015reliability,balajical19,das2013aging,das2013reliability,das2014combined,das2014communication,das2012fault,SrinivasanISCA04,das2014energy}. We illustrate NBTI, which is a dominant reliability issue in scaled technology nodes. NBTI-induced aging of a CMOS device in a peripheral circuit at temperature \ineq{T} is calculated as 
\begin{equation}
    \label{eq:nbti_aging}
    \footnotesize \mathcal{A}(T) = \sum_{i=0}^{N_a-1} g_0(T)\cdot V_{\text{bias}}^a\cdot tRC^b,
\end{equation}
where \ineq{N_a} is the number of PCM accesses, \ineq{g_0(T)}, \ineq{a}, and \ineq{b} are material-dependent constants~\cite{balajical19}. The bias voltage for a specific PCM operation, \ineq{V_{\text{bias}}}, is obtained from Table~\ref{tab:bias}, and the PCM timing parameter, \ineq{tRC}, from Table~\ref{tab:min_max_latency}.

Using Equation~\ref{eq:nbti_aging}, the reliability (\ineq{R(T)}) and lifetime (\ineq{L_a}) can be computed as
\begin{equation}
    \label{eq:reliability_lifetime_nbti}
    \footnotesize R(T) = e^{-\mathcal{A}(T)^\beta} \text{ and } L_a = \int R(T).
\end{equation}
From Equations~\ref{eq:nbti_aging} and \ref{eq:reliability_lifetime_nbti}, we can conclude that aging can be used as a measure of lifetime.
In Section~\ref{sec:results}, we show improvements of \tech{} in terms of \ineq{L_e} and \ineq{\mathcal{A}}.

\subsection{Efficient Implementation of FTI Table and AIR}
\label{sec:storing}
We now discuss an efficient implementation of \tech{}.

\vspace{10pt}

\subsubsection{Implementing FTI Table}
\tech{} stores most acc\-ess-inducing FTIs (i.e., their program counters) from an execution phase in the FTI Table. A naive way to implement the FTI table is to use a row for each FTI. However, the exact number of rows within this table will vary depending on the number of FTIs, which is program-specific. If a table's capacity is inadequate to store all high-access inducing FTIs, \tech{} will start predicting every new FTI as non access-intensive, once the table is full. The OS page fault handler will then allocate all new pages to the far memory segment, providing no significant performance improvement. 
Therefore, the FTI table must be sized conservatively (i.e. assuming the maximum number of FTIs), leading to a large hardware cost for table storage and lookup. We propose to use Bloom filer to implement the FTI table.


The \emph{Bloom filter} is a memory-efficient data structure to represent set membership~\cite{bloom1970space}. The price paid for this efficiency is that this filter is a probabilistic data structure: it tells if an element is either \textit{definitely not} in the set (zero false positive) or \textit{may be} in the set (non-zero false negative). 

\begin{figure}[h!]
	\centering
	\centerline{\includegraphics[width=0.79\columnwidth]{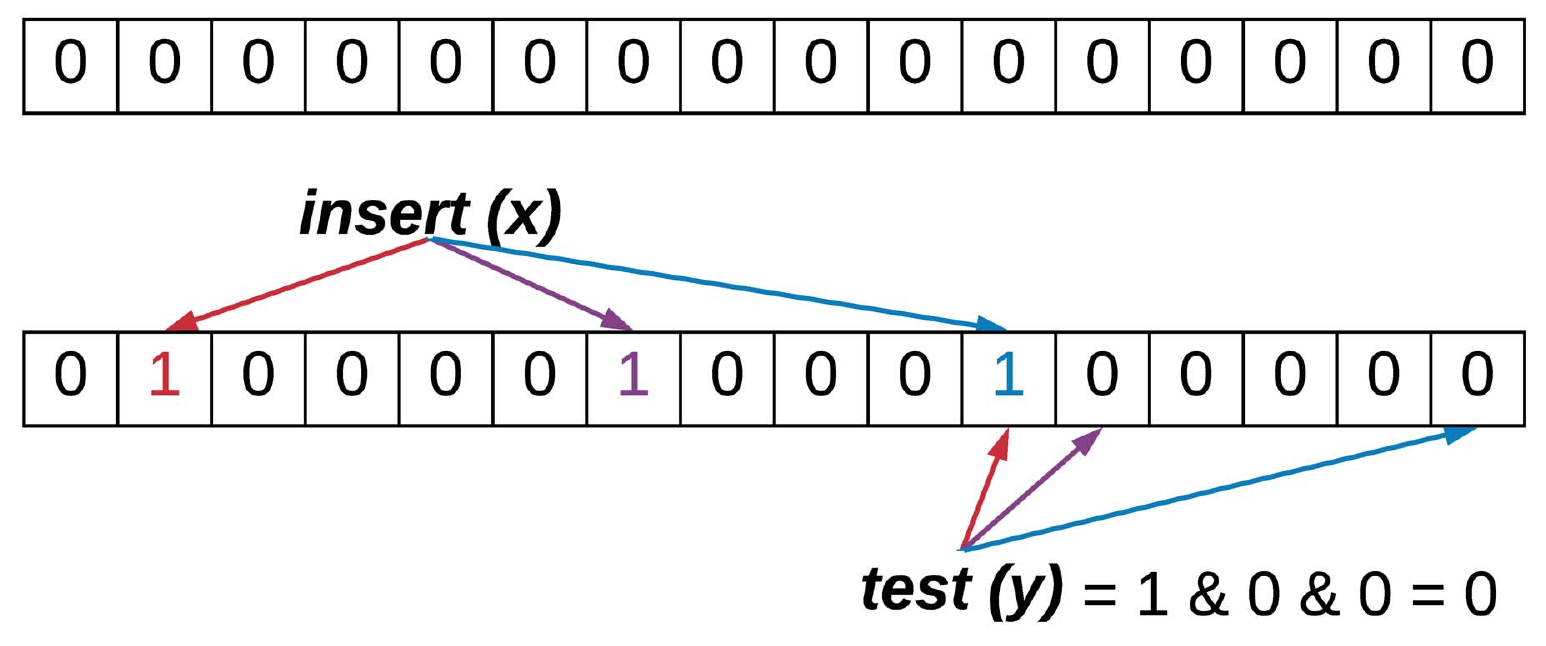}}
	\vspace{-10pt}
	\caption{Operation of a Bloom filter.}
	\vspace{-10pt}
	\label{fig:bloom}
\end{figure}

The filter is implemented as a bit array of length \ineq{m} with \ineq{k} distinct hash functions. Figure~\ref{fig:bloom} shows a filter in which \ineq{m = 16} and \ineq{k = 3}. 
To insert an element \ineq{x} in the filter, it is hashed with all three functions, and all of the bits in the corresponding positions are set to 1. 
Conversely, to test if an element \ineq{y} is in the filter, it is again hashed using all three functions; and if all of the bits at the corresponding bit positions are 1, the element is declared to be present in the Bloom filter. In Fig.\ref{fig:bloom}, \ineq{y} is declared to be not present.

Because bits are never reset: 1) an element once inserted cannot be deleted from the filter\footnote{Certain modifications to the Bloom filter allow for deletion of elements. One example is the Cuckoo filter \cite{fan2014cuckoo}.}, and 2) a false negative can never occur. The rate of false positive is approximately 
\begin{equation}
    \label{eq:false_positive}
    \footnotesize \left(1-e^{-kn/m}\right)^k,
\end{equation}
where \ineq{n} is the number of elements expected to be inserted in the filter. \tech{} uses two Bloom filters for the FTI\_R and FTI\_W tables, each implemented with \ineq{m = 128}, \ineq{k = 3}.

\vspace{10pt}

\subsubsection{AIR Implementation}
The AIR is implemented as a table with \ineq{D} rows, each having five fields: a valid field, the program counter value of the FTI, the number of read and write inducing pages touched by this FTI, and the number of accesses that go to these pages.
Within a program phase, the least-frequently used entry in the AIR is overwritten with a new FTI; the field recording the total number of accesses is used as the frequency estimate. Upon completion of a phase, any entry having frequency higher than a threshold is inserted into the Bloom filter. Subsequently, the AIR is reset by setting the valid field of all its entries to 0.

%% file: sections/evaluation.tex
To evaluate \tech{}, we develop a cycle-accurate DRAM-PCM hybrid memory simulator with the following:
\begin{itemize}
	\item A Cycle-level x86 multi-core simulator, whose front-end  is based on Pin~\cite{LukPin}. We configure this to simulate 8 out-of-order cores. 
	\item A main memory simulator, closely matching the JEDEC Nonvolatile Dual In-line Memory Module (NVDIMM)-N/F/P Specifications~\cite{jedecnvdimm2017}. This simulator is composed of Ramulator~\cite{kim2016ramulator}, to simulate DRAM, and a cycle-level PCM simulator based on NVMain~\cite{poremba2015nvmain}.
	\item Power and latency for DRAM and PCM are based on Intel/Micron's 3D Xpoint specification~\cite{bourzac2017has,pcm_book}. Energy is modeled for DRAM using DRAMPower~\cite{chandrasekar2012drampower} and for NVM using NVMain with parameters from~\cite{pcm_book}.
\end{itemize}

Table~\ref{tab:simulation_parameters} summarizes the various simulation parameters.

\begin{table}[h!]
\setlength{\tabcolsep}{5pt}
\renewcommand{\arraystretch}{1.0}
\caption{Major simulation parameters.}
\label{-10pt}
\label{tab:simulation_parameters}
\centering
{\fontsize{8}{10}\selectfont
\begin{tabu}{r p{5.5cm}}
    \tabucline[2pt]{-}
    Processor & 8 cores, 3 GHz, out-of-order\\
    L1-I/D cache & Private 64KB per core, 4-way\\
    L2 cache & shared, 4MB, 8-way \\
    \hline
    DRAM Main Memory & 64GB, Micron DDR3\\
    & 2 channels, 4 ranks/channel, 8 banks/rank, 128 sub-arrays/bank, 512 rows/sub-array\\
    & Memory clock = 1066MHz\\
    & Near bitline segment = 128 cells\\
    & Far bitline segment  = 384 cells (= 512 -128)\\
    \hline
    PCM Main Memory & 128GB, Micron DDR3 \cite{pcm_book}\\
    & 4 channels, 4 ranks/channel, 8 banks/rank, 8 partitions/bank, 128 tiles/partition, 4096 rows/tile\\
    & Memory clock = 1066MHz\\
    & Near bitline segment = 512 cells\\
    & Far bitline segment  = 3584 cells(= 4096 -512)\\
    \tabucline[2pt]{-}
\end{tabu}
}
\end{table}


We evaluate the following main memory architectures.
\begin{itemize}
    \item \textbf{M1:} DRAM-PCM hybrid memory with DRAM and PCM placed on separate DIMMs, sharing a common main memory address space. This is the primary hybrid memory architecture that we evaluate in this paper, and is similar to Intel Optane~\cite{bourzac2017has} using PCM instead of SSD.
    \item \textbf{M2:} PCM main memory with DRAM as write cache. This is similar to the architecture of IBM Power9~\cite{sadasivam2017ibm}.
    \item \textbf{M3:} Mainstream DRAM-based main memory architecture similar to Intel Skylake~\cite{doweck2017inside}.
\end{itemize}

We evaluate architectures M2 and M3 to show that \tech{} improves performance of other memory architectures.

We evaluate the following techniques.
\begin{itemize}
    \item \textit{Baseline} allocates a page randomly to a free physical address. Pages are not migrated between DRAM and PCM during program execution. Bitlines are not segmented. 
    \begin{itemize}
        \item for M1, the Baseline is Intel Optane~\cite{bourzac2017has}.
        \item for M2, the Baseline is IBM Power9~\cite{sadasivam2017ibm}.
        \item for M3, the Baseline is Intel Skylake~\cite{doweck2017inside}.
    \end{itemize}
    \item \textit{Nimble}~\cite{yan2019nimble} supports M1-type hybrid memory. It  migrates pages between DRAM and PCM during program execution, starting from a random physical address allocation. Bitlines 
    are not segmented.
    \item \textit{TL-DRAM}~\cite{lee2013tiered} supports DRAM (M3). It  uses the page management policy of Baseline. Each bitline in a DRAM bank is partitioned into near and far segments.
    \item \textit{\tech{}} supports both M1 and M2-type hybrid memory architectures. It  1) uses segmented bitlines for DRAM and PCM, 2) controls initial page allocation to correct memory tiers, and 3) minimizes channel occupancy during page migrations between tiers of the same memory, to improve performance and reliability.
\end{itemize}

\begin{figure}[h!]
	\centering
	\centerline{\includegraphics[width=0.99\columnwidth]{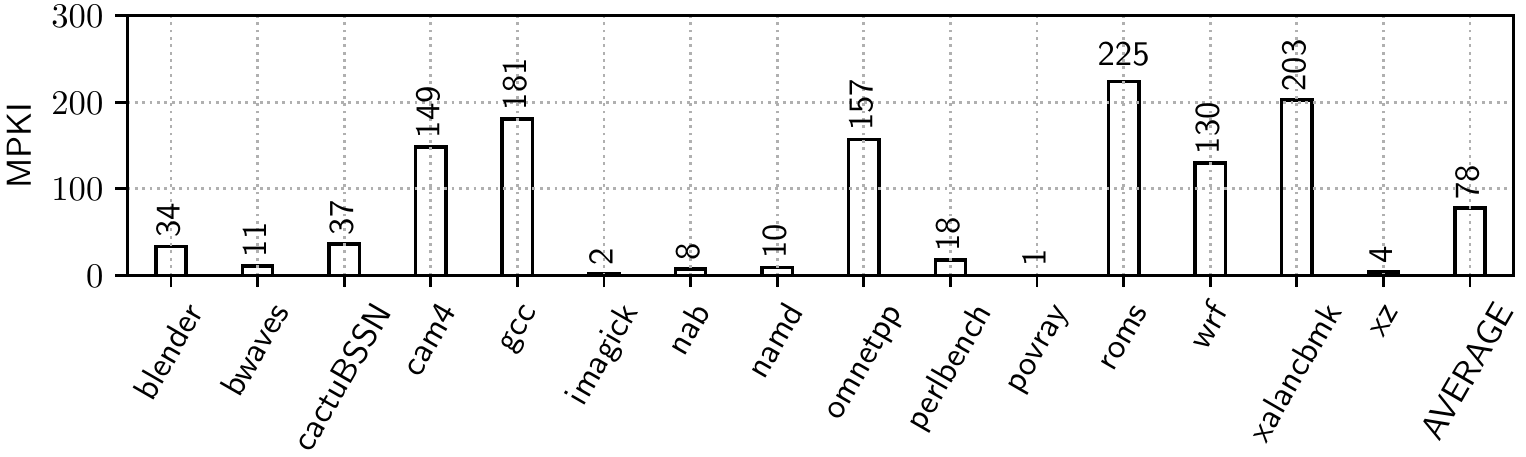}}
	\vspace{-10pt}
	\caption{MPKI for our evaluated workloads.}
	\vspace{-10pt}
	\label{fig:mpki}
\end{figure}

We evaluate all single-core and multi-programmed workloads from the SPEC CPU2017 suite~\cite{bucek2018spec}.
Table~\ref{tab:spec} reports the workloads that we present in Section~\ref{sec:results}. These workloads are chosen because they have at least 1 cache Miss Per Kilo Instructions (MPKI) (see Fig.~\ref{fig:mpki}). For other workloads with low MPKI (those not presented in Sec.~\ref{sec:results}), \tech{} neither significantly improves nor hurts performance and reliability.

\begin{table}[h!]
\setlength{\tabcolsep}{2pt}
\renewcommand{\arraystretch}{1.0}
\caption{Evaluated workloads.}
\vspace{-5pt}
\label{tab:spec}
\centering
{\fontsize{8}{10}\selectfont
\begin{tabu}{r p{6cm}}
    \tabucline[2pt]{-}
    single-core & 8 copies each of blender, bwaves, cactuBSSN, cam4, gcc, imagick, nab, namd, omnetpp, perlbench, povray, roms, wrf, xalancbmk, xz\\
    multi-programmed & \textbf{MP1} (2 copies each of blender, bwaves, cactuBSSN, and cam4) and \textbf{MP2} (2 copies each of perlbench, wrf, xalancbmk, and xz)\\
    \tabucline[2pt]{-}
\end{tabu}
}
\end{table}


All workloads are executed for 10 billion instructions.

%% file: sections/results.tex
\subsection{Summary of Key Results}
\label{sec:summary}
Table \ref{tab:compare_sota} summarizes \tech{}'s improvements.


\begin{table}[h!]
\setlength{\tabcolsep}{3pt}
\renewcommand{\arraystretch}{1.0}
\caption{Summary of key results.}
\label{tab:compare_sota}
\centering
{\fontsize{8}{10}\selectfont
\begin{tabu}{r | c c c c c}
    \tabucline[2pt]{-}
    & System & Energy & Migration & \multicolumn{2}{c}{Lifetime}\\
    & Perf. & Consump. & Overhead & \multicolumn{2}{c}{(Sec. \ref{sec:reliability})}\\\cline{5-6}
    \tech{} vs. & (Sec. \ref{sec:perf}) & (Sec. \ref{sec:energy}) & (Sec. \ref{sec:overheads}) & Endurance & Aging\\
    \hline
    Intel Optane & $21\%\uparrow$ & $19\%\downarrow$ & -- & -- & --\\
	Nimble \cite{yan2019nimble} & $16\%\uparrow$ & $18\%\downarrow$ & $71.2\%\downarrow$ & $20\%\uparrow$ & $33\%\downarrow$\\\cline{3-6}
	IBM Power9 & $15\%\uparrow$ & \multicolumn{4}{|c}{~}\\
	Intel Skylake & $15\%\uparrow$ & \multicolumn{4}{|c}{~}\\
	TL-DRAM \cite{lee2013tiered} & $13\%\uparrow$ & \multicolumn{4}{|c}{~}\\
	\hline
    \tabucline[2pt]{-}
\end{tabu}
}
\end{table}

\subsection{Overall System Performance}
\label{sec:perf}
We report overall system performance for three configurations: 1) \emph{M1:} DRAM-PCM hybrid memory with DRAM and PCM on separate DIMMs, 2) \emph{M2:} DRAM-PCM hybrid memory with DRAM as write cache to PCM, and 3) \emph{M3:} DRAM-based system.


\subsubsection{Hybrid Main Memory Architecture M1}
\begin{figure}[h!]
	\centering
	\centerline{\includegraphics[width=0.99\columnwidth]{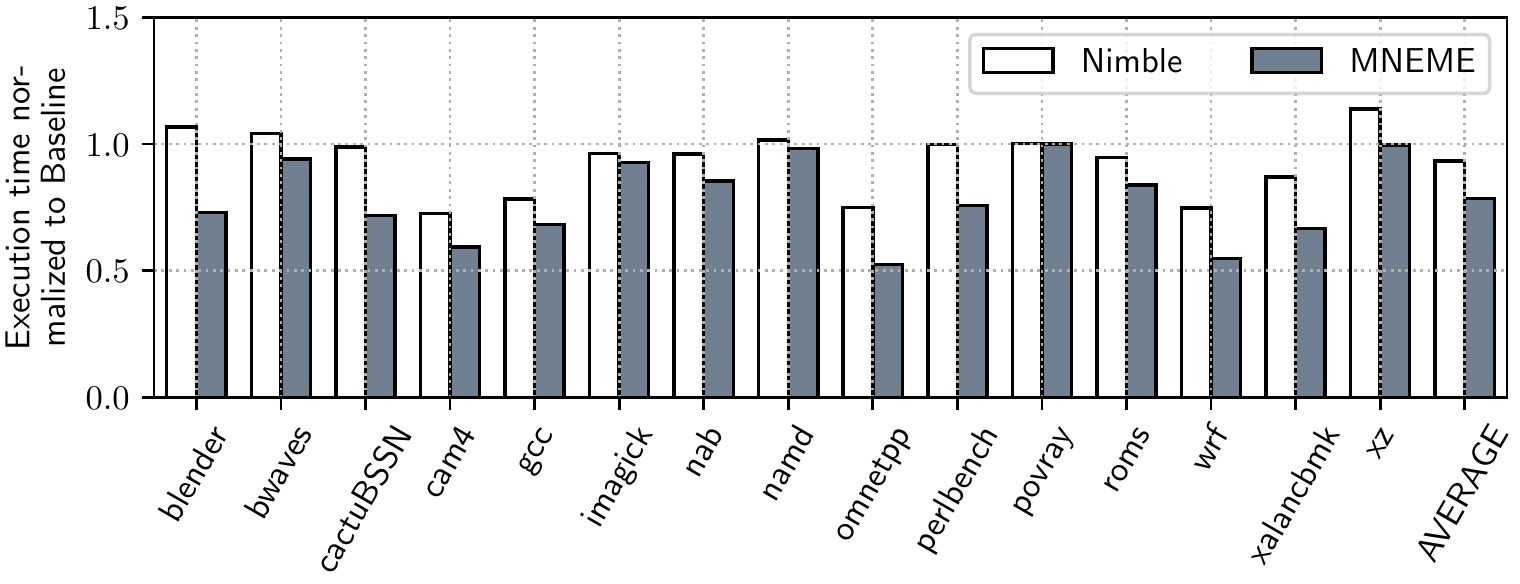}}
	\vspace{-5pt}
	\caption{Execution time, normalized to Baseline for DRAM-PCM hybrid main memory with shared address.}
	\vspace{-5pt}
	\label{fig:m1}
\end{figure}

Figure~\ref{fig:m1} reports the execution time of each workload for our evaluated systems normalized to Baseline. The simulator is configured for our primary DRAM-PCM hybrid main memory architecture with a common address space. We make the following three main observations.

First, Nimble achieves better performance than Baseline by an average of 7\% due to Nimble's policy to migrate hot pages from PCM to DRAM, which reduces execution time (DRAM has lower access latency than PCM). Second, for workloads such as blender and bwaves, performance of Nimble is, in fact, worse than Baseline because of the high overhead of page migrations in Nimble. Third, \tech{}'s performance is the best among all three systems. On average, the execution time of \tech{} is 21\% lower than Baseline and 16\% lower than Nimble. This improvement is due to 1) \tech{}'s segmented bitline architecture and 2) \tech{}'s intelligent initial page allocation policy to exploit performance asymmetries in memory tiers.


\subsubsection{Hybrid Main Memory Architecture M2}
\begin{figure}[h!]
	\centering
	\centerline{\includegraphics[width=0.99\columnwidth]{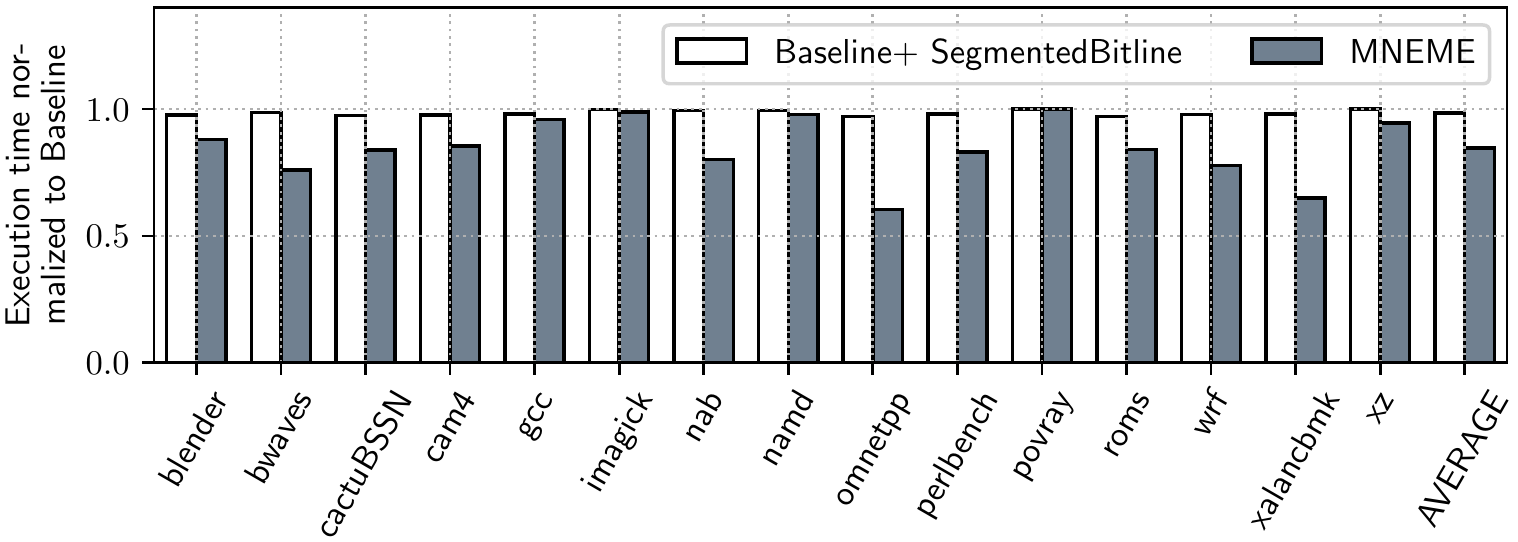}}
	\vspace{-10pt}
	\caption{Execution time, normalized to Baseline for DRAM-PCM hybrid memory with DRAM as write cache.}
	\vspace{-5pt}
	\label{fig:m2}
\end{figure}

Figure~\ref{fig:m2} reports the execution time of each workload for our evaluated systems normalized to Baseline with the simulator configured for DRAM-PCM hybrid main memory with DRAM configured as write cache to PCM. We make the following two main observations.

First, using the proposed segmented bitline architecture, performance of Baseline improves only marginally, by an average of 2\% (first bar in each set). See also Observation 1. Second, \tech{}'s performance is the highest among all three systems. On average, \tech{}'s execution time is 15\% lower than Baseline and 14\% lower than segmented bitlines. 


\subsubsection{DRAM-based Main Memory Architecture M3}
\begin{figure}[h!]
	\centering
	\centerline{\includegraphics[width=0.99\columnwidth]{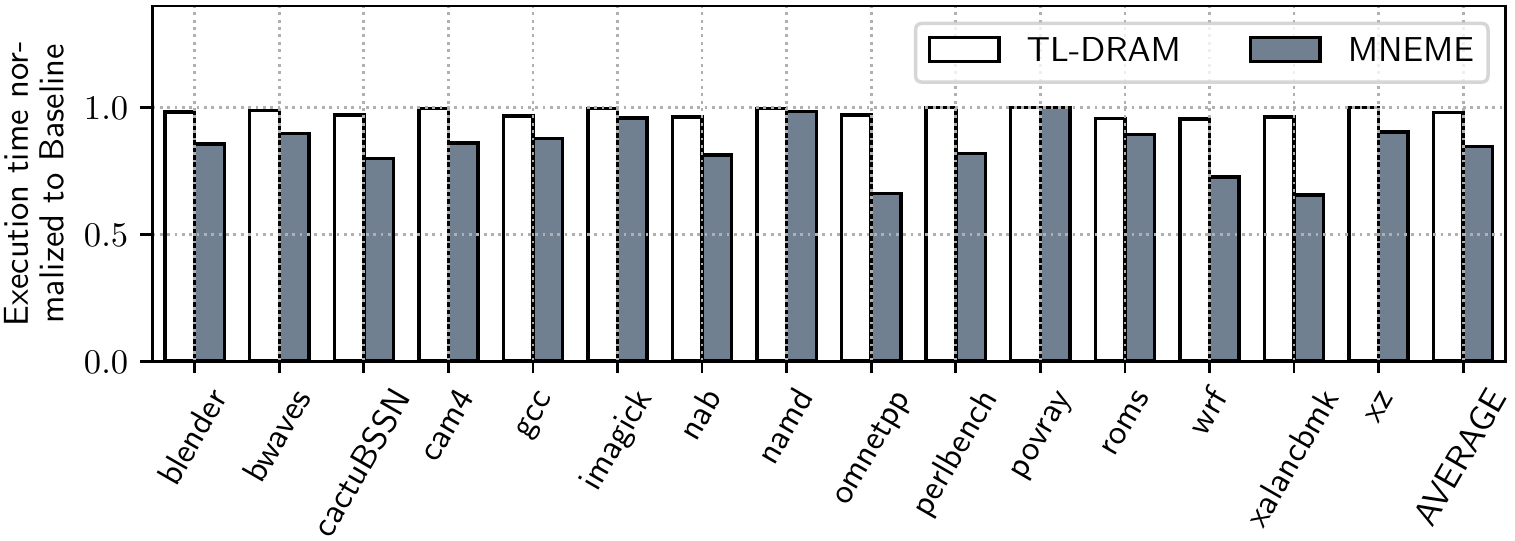}}
	\vspace{-10pt}
	\caption{Execution time, normalized to Baseline for DRAM-based main memory.}
	\vspace{-10pt}
	\label{fig:m3}
\end{figure}

~\\
Figure~\ref{fig:m3} reports the execution time of each workload for our evaluated systems normalized to Baseline with the simulator configured for DRAM-based main memory. We observe that performance of TL-DRAM is marginally better than Baseline. \tech{} has the highest performance among all three systems. On average, \tech{}'s execution time is 15\% lower than Baseline and 13\% lower than TL-DRAM.

Unless otherwise stated, following results are for our primary memory architecture, i.e., DRAM-PCM hybrid memory with a common memory address space (M1).

\subsection{Multi-Programmed Workloads}
\label{sec:multi_programmed}
\begin{figure}[h!]
	\centering
	\centerline{\includegraphics[width=0.99\columnwidth]{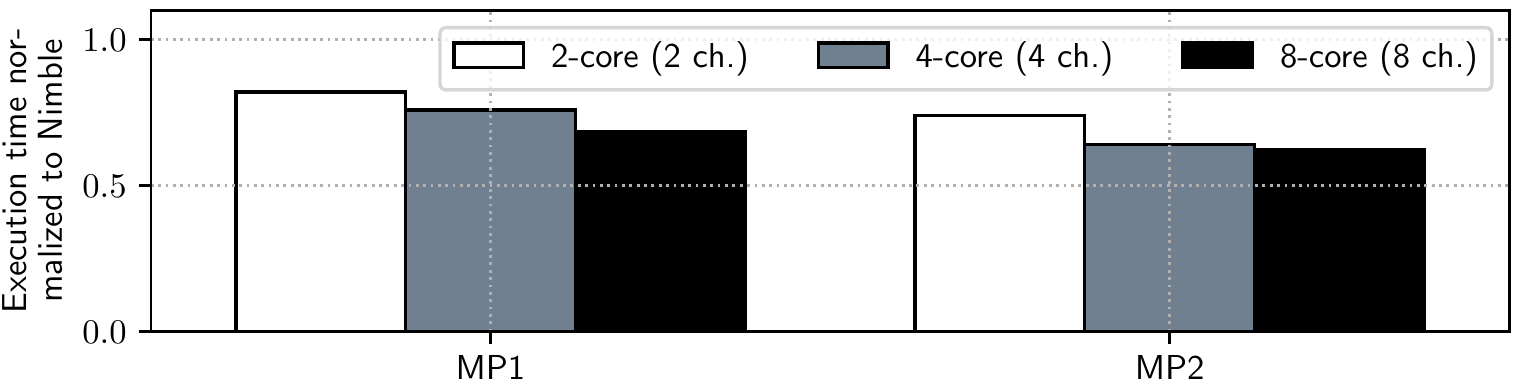}}
	\vspace{-10pt}
	\caption{Execution time normalized to Nimble for multi-programmed workloads on 2-core, 4-core, and 8-core.}
	\vspace{-10pt}
	\label{fig:varying_cores}
\end{figure}

Figure \ref{fig:varying_cores} plots the execution time of \tech{} normalized to Nimble for 2 multi-programmed workloads on 2-core (2-channel), 4-core (4-channel), and 8-core (8-channel) systems.

We observe that for 2, 4, and 8 cores in the system, \tech{} provides 18\%, 24\% and 31\% performance improvement for MP1, and 26\%, 36\% and 38\% performance improvement for MP2, compared to Nimble. 
The performance improvement of \tech{} increases with increasing
number of channels. This is because, with more channels,
bank
access latency becomes the primary performance bottleneck. Therefore, \tech{},
which reduces the average bank access latency, provides
better performance with more channels.

\subsection{Memory Access Distribution}
\label{sec:access}
\begin{figure}[h!]
	\centering
	\centerline{\includegraphics[width=0.99\columnwidth]{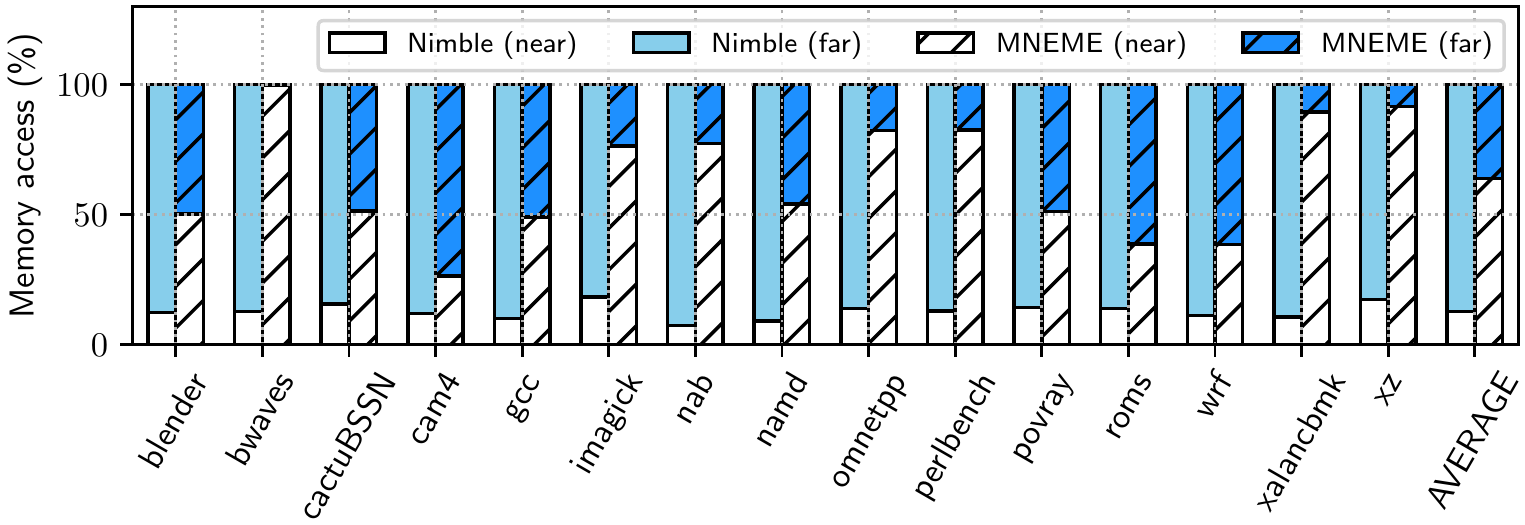}}
	\vspace{-10pt}
	\caption{Memory accesses to near and far segments.}
	\vspace{-10pt}
	\label{fig:access_distribution}
\end{figure}

Figure~\ref{fig:access_distribution} plots the memory accesses to near and far memory segments of each workload for Nimble and \tech{}. 
We make the following two main observations.


First, on average, only 13\% of accesses go to near segments in memory banks using the initial page allocation and hot page migration policy of Nimble. 
Second, 
\tech{} directs an average of 64\% of accesses to near memory segments using its intelligent initial page allocation policy. This leads to significant performance improvement (see Section~\ref{sec:perf}).

\subsection{Migration Overhead}
\label{sec:overheads}
\begin{figure}[h!]
	\centering
	\centerline{\includegraphics[width=0.99\columnwidth]{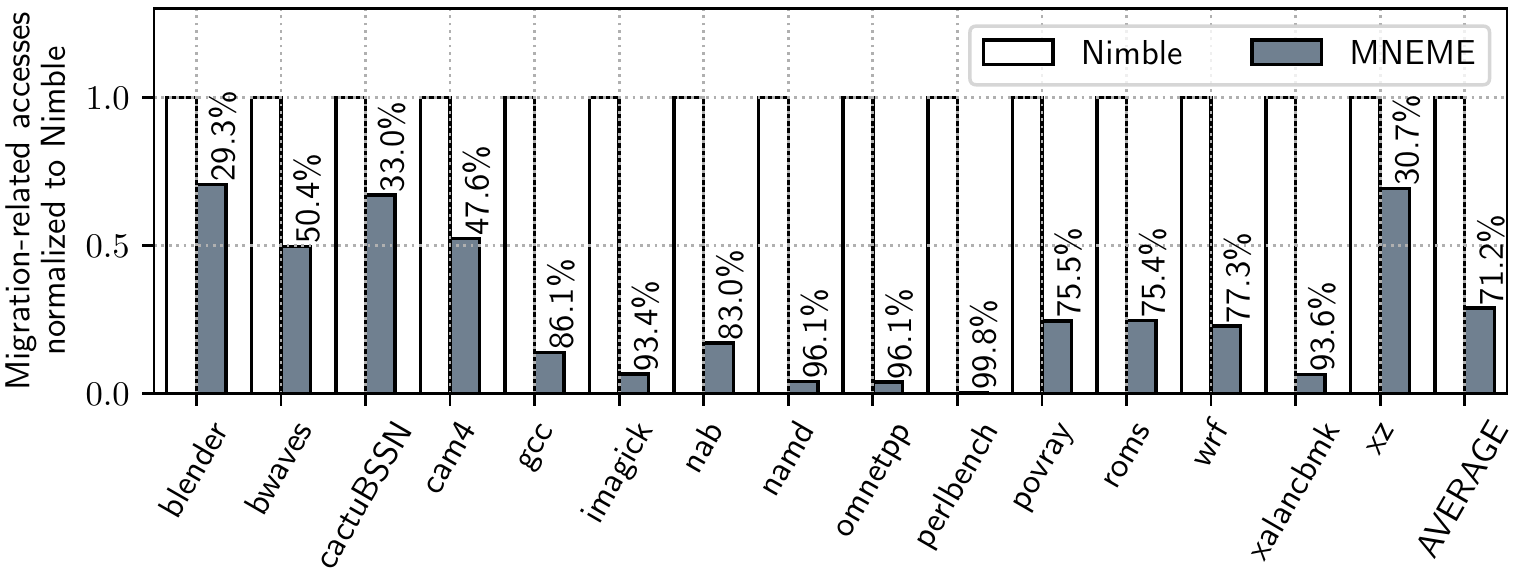}}
	\vspace{-10pt}
	\caption{Migration-related accesses normalized to Nimble.}
	\vspace{-10pt}
	\label{fig:migration_overhead}
\end{figure}

Figure \ref{fig:migration_overhead} plots the page migration-related accesses in \tech{} normalized to Nimble for each workload. We observe that page migration-related accesses in \tech{} are lower than Nimble by an average of 71.2\%. This reduction is because 1) \tech{} places new pages in correct memory tiers during their initial allocation using access profiles of observed first-touch instructions in the program, which reduces the average number of inter-memory page migrations, and 2) \tech{} uses its new peripheral circuit design to eliminate channel usage for page migrations within each memory bank. 

\subsection{Length of Execution Phases}
\label{sec:interval}
\begin{figure}[h!]
	\centering
	\centerline{\includegraphics[width=0.99\columnwidth]{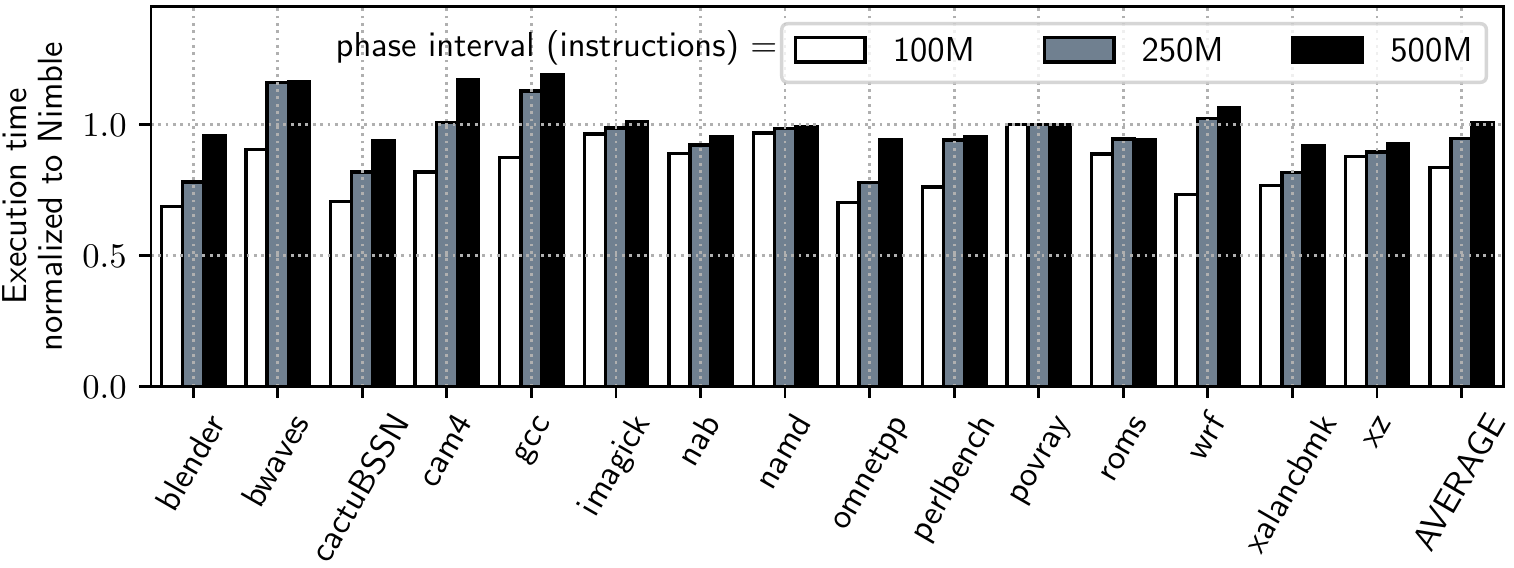}}
	\vspace{-10pt}
	\caption{Execution time, normalized to Nimble for program phase intervals of 100M, 250M, and 500M instructions.}
	\vspace{-10pt}
	\label{fig:interval}
\end{figure}

Figure~\ref{fig:interval} reports the execution time of \tech{} normalized to Nimble for each workload. The first bar in each set is for the default phase length of 100 million instructions. The second and third bars are for phase length of 250 million and 500 million instructions, respectively.
We observe the execution time of \tech{} to increase with the length of the phase interval. This is because, lower phase intervals allow finer control of page allocation, resulting in higher performance (i.e., lower execution time) than Nimble. However, lower phase intervals also result in higher overhead due to 1) frequent updates to FTI Table and AIR and 2) frequent page migrations and updates to page table, impacting performance. Phase interval of 100 million instructions gives the best performance and overhead trade-off.

\subsection{FTI-based Access Intensity Prediction}
\label{sec:adaptive_improvement}
\begin{figure}[h!]
	\centering
	\centerline{\includegraphics[width=0.99\columnwidth]{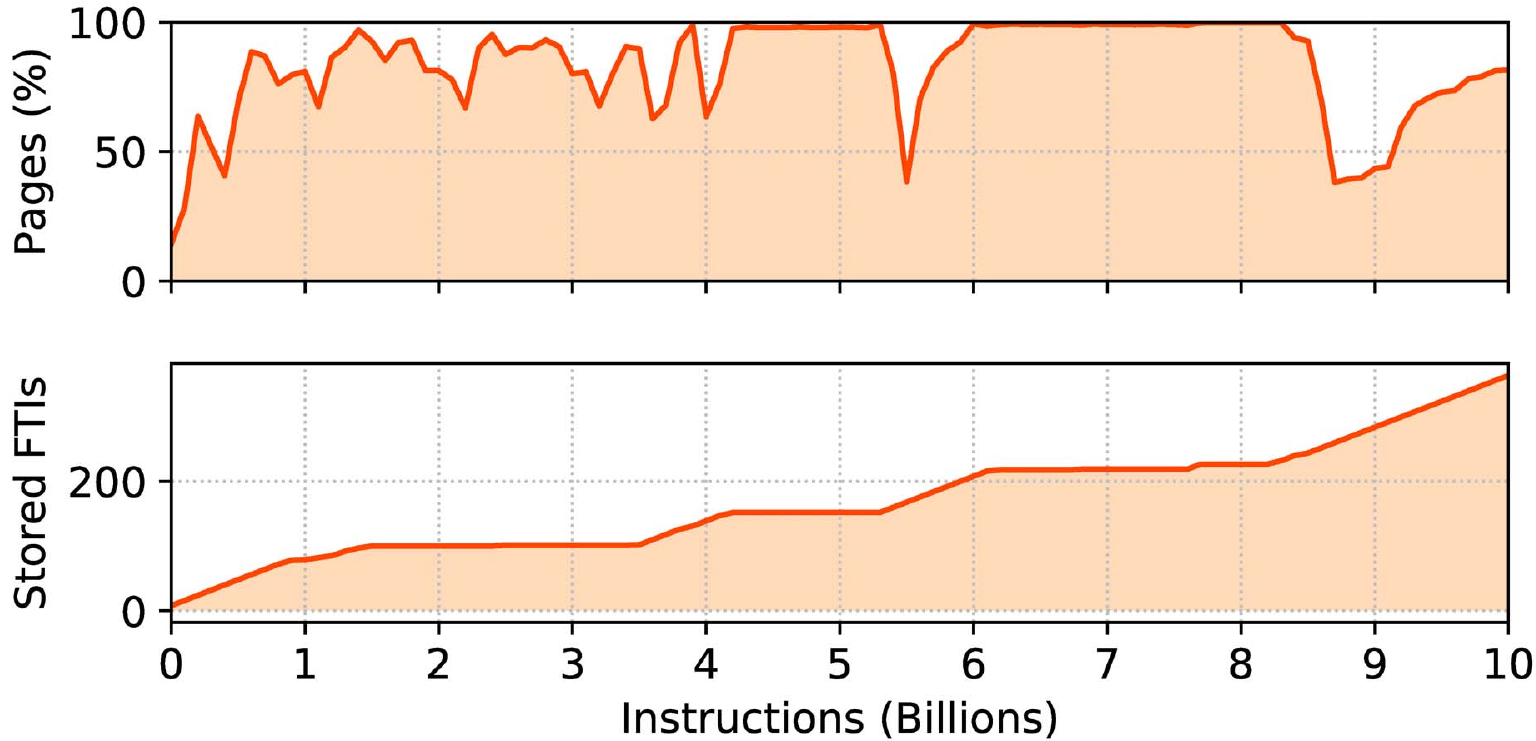}}
	\vspace{-10pt}
	\caption{Illustration of the FTI-based access intensity prediction for cam4.}
	\vspace{-10pt}
	\label{fig:adaptive}
\end{figure}

Figure \ref{fig:adaptive} illustrates how \tech{} improves its page allocation decisions over time, 
improving overall performance. The bottom subfigure shows the increase in stored FTIs during the execution of cam4. The top subfigure shows the fraction of memory pages (i.e., their program counters) that hit in the FTI Table. 
We observe that cam4 undergoes a change in behavior after executing \ineq{\approx 5.5} billion instructions and then again after \ineq{\approx 8.7} billion instructions, due to potentially distinct work sets. We see a dip in the number of pages that hit in the FTI Table (top subfigure). Therefore, the number of stored FTIs increases sharply around these time (bottom subfigure) because \tech{} starts inserting the newly observed FTIs into the FTI table to improve its allocation decisions. This results in an increase in the number of page hits (top subfigure) during subsequent execution of cam4.

\subsection{Energy Consumption}
\label{sec:energy}
\begin{figure}[h!]
	\centering
	\centerline{\includegraphics[width=0.99\columnwidth]{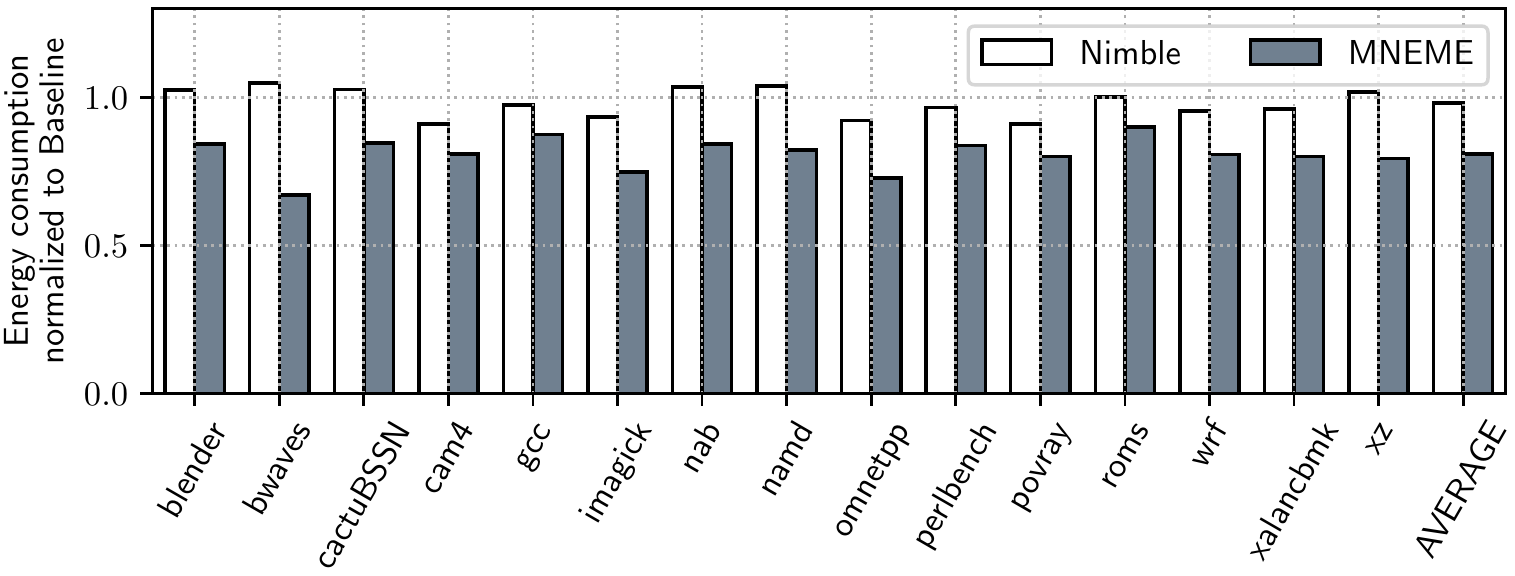}}
	\vspace{-10pt}
	\caption{Energy consumption, normalized to Baseline.}
	\vspace{-10pt}
	\label{fig:energy}
\end{figure}

Figure~\ref{fig:energy} reports the total energy consumption (demand accesses and page migrations) of each workload for our evaluated systems normalized to Baseline. We observe that Nimble has lower energy consumption than Baseline by an average of only 2\%. Although energy consumption of demand accesses is lower in Nimble, the potential energy savings are overshadowed by the page migrations.
\tech{} has the lowest energy consumption (on average, 19\% lower than Baseline and 18\% lower than Nimble). These savings are achieved in \tech{} because it reduces page migrations on the memory channel significantly by 1) initially allocating a page to a correct tier, and 2) facilitating inter-segment data transfers, which reduce channel occupancy.

\subsection{Reliability}
\label{sec:reliability}
We evaluate two reliability issues: endurance and NBTI.

\vspace{10pt}

\subsubsection{Endurance-related Lifetime}
\begin{figure}[h!]
	\centering
	\centerline{\includegraphics[width=0.99\columnwidth]{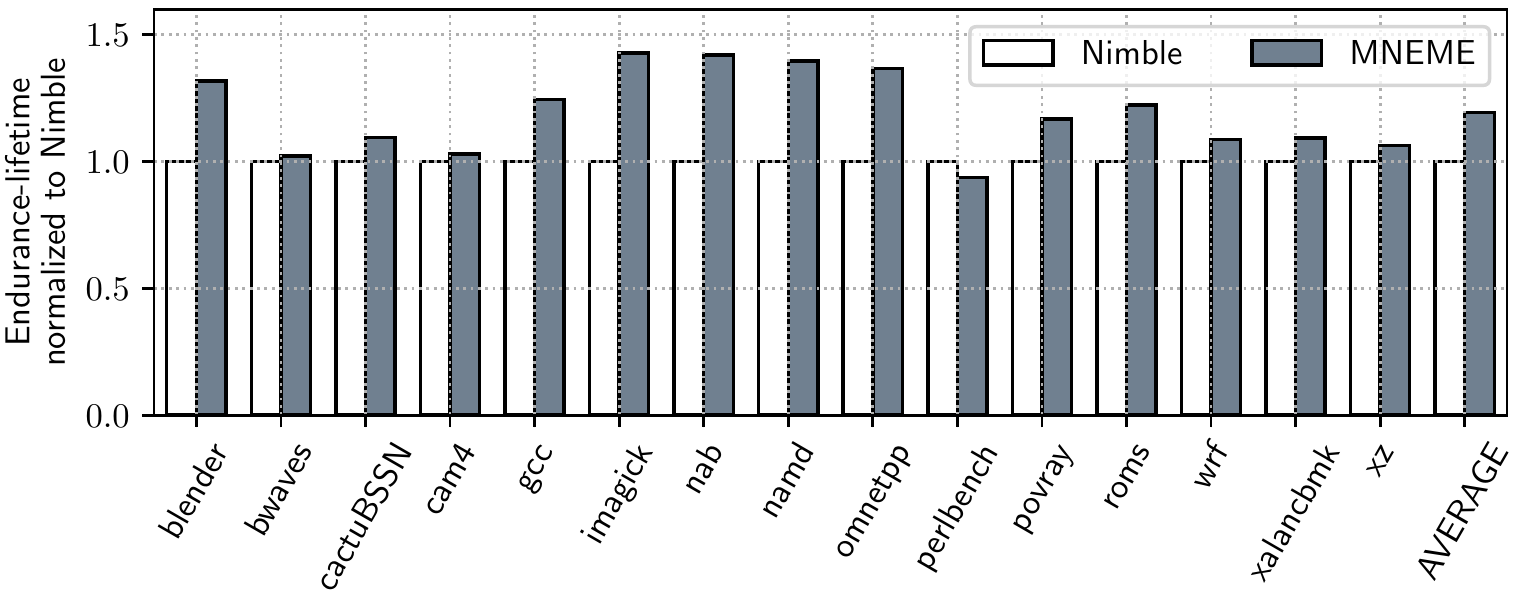}}
	\vspace{-10pt}
	\caption{Endurance lifetime, normalized to Nimble.}
	\vspace{-10pt}
	\label{fig:endurance}
\end{figure}

Figure~\ref{fig:endurance} reports the endurance-related lifetime (computed using Equation~\ref{eq:endurance_lifetime}) of each workload for our evaluated systems normalized to Nimble. We make the following observation.

\tech{} improves endurance-related lifetime by an average of 20\% compared to Nimble. This improvement is because of the extra tiers that \tech{} creates inside each memory using isolation transistors. Pages that are not frequently referenced can now stay in DRAM far segments for sometime, before they are migrated to PCM. This reduces PCM writes, which improves endurance. 



\subsubsection{NBTI-related Aging}
\begin{figure}[h!]
	\centering
	\centerline{\includegraphics[width=0.99\columnwidth]{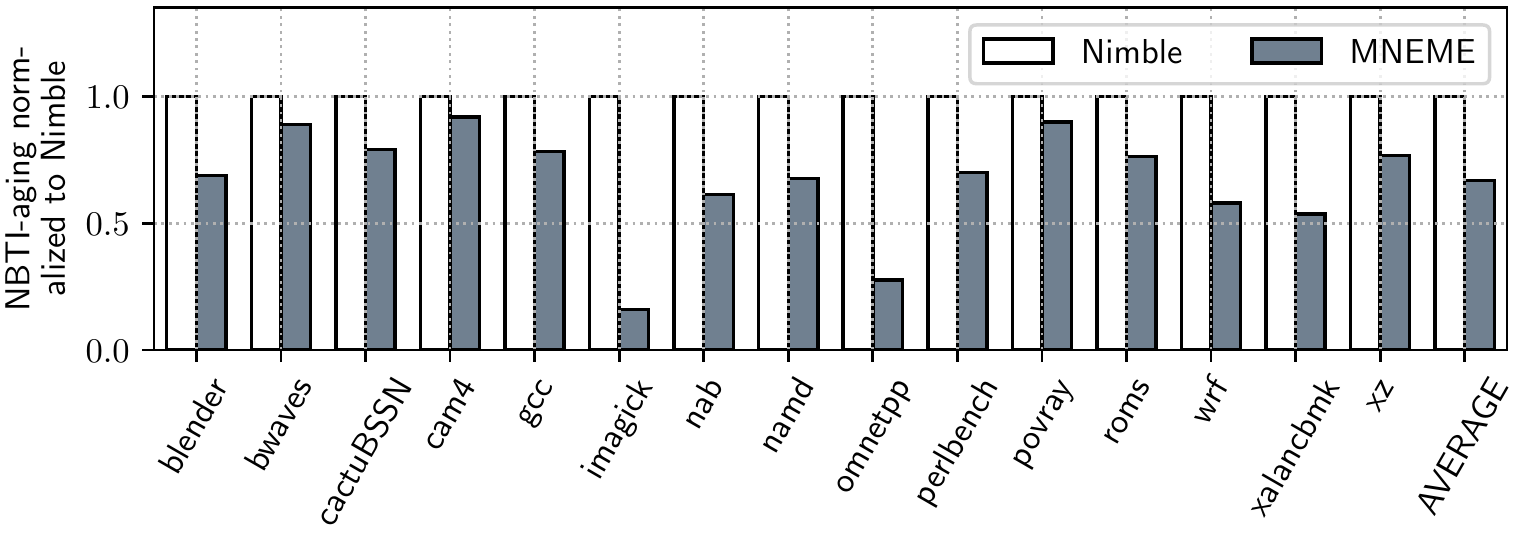}}
	\vspace{-10pt}
	\caption{NBTI-related aging, normalized to Nimble.}
	\vspace{-10pt}
	\label{fig:aging}
\end{figure}

Figure~\ref{fig:aging} reports the NBTI-related aging (computed using Equation~\ref{eq:nbti_aging}) of each workload for our evaluated systems normalized to Nimble. We observe that
\tech{} has 33\% lower NBTI-related aging than Nimble. This is because \tech{} uses lower bias voltages to access near PCM segments (Table \ref{tab:bias}) due to its segmented bitline architecture. Also, the CMOS devices in PCM's peripheral circuit are stressed for reduced time duration than Nimble due to lower timing requirements of the near segments (Table \ref{tab:min_max_latency}). Both these factors contribute to lower NBTI aging~\cite{balajical19}.


\subsection{Design Area Analysis}
\tech{} introduces two changes: increase in memory die size due to isolation transistors and overhead in the memory controller to maintain FTI Table and AIR.


\subsubsection{Area Overhead on Memory-side}~\\
Adding an isolation transistor to each bitline increases the area of each bank. We estimate this for DRAM and PCM as follows. 
In DRAM, the sense amplifier and the isolation transistor are respectively 115.2x and 11.5x taller than an individual DRAM cell. For a subarray of 512 DRAM cells per bitline, the area overhead is \ineq{\frac{11.5}{115.2 + 512} = 1.83\%}~\cite{lee2013tiered}. 

In PCM, the peripheral circuit and the isolation transistor are respectively 384x and 9.6x taller than an individual PCM cell. For a PCM partition of 4096 PCM cells per bitline, the area overhead is \ineq{\frac{9.6 + 9.6}{384 + 4096} = 0.43\%} (including the change to support in-memory migrations).


\subsubsection{Area Overhead on CPU-side}~\\
FTI Tables are implemented as two 128-bit Bloom filters with a total size of 256 bits. AIR is implemented as an 8-entry table with 1-bit valid field, a 32-bit field for the program counter, a 32-bit field for counting accesses, and two 16-bit fields for counting pages. The total area overhead of AIR is 97B. The LUT stores 4 extra rows in Table~\ref{tab:min_max_latency} (2 for PCM read and write latency of near segment and 2 for DRAM read and write latency of near segment). The extra area overhead is 160 bits (= 4 * 5 entries per row * 8 bits per entry). So, \tech{} introduces a total of 149 bytes in storage in the memory controller, corresponding to an area overhead of \ineq{6\times 10^{-4}\text{mm}^2} at 45nm. 
Given the cost sensitivity of memory designs, designers can still benefit from \tech{}'s standalone page allocation policy, without segmented bitlines. 

\begin{figure}[h!]
	\centering
	\centerline{\includegraphics[width=0.99\columnwidth]{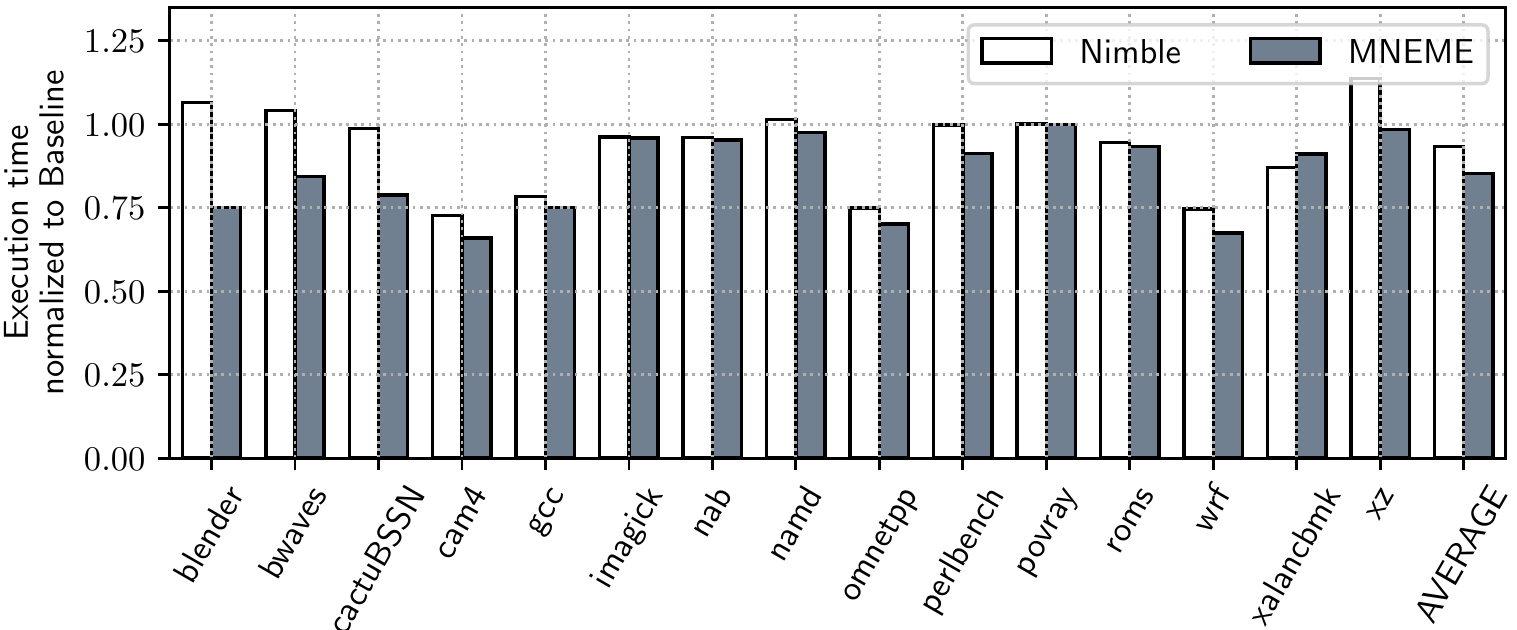}}
	\vspace{-10pt}
	\caption{Execution time, normalized to Baseline for mainstream memory without segmented bitlines.}
	\vspace{-10pt}
	\label{fig:no_tiling}
\end{figure}

Figure~\ref{fig:no_tiling} plots the execution time of each workload for Nimble and \tech{}, normalized to Baseline. 
The simulator is configured for DRAM-PCM hybrid memory architecture with shared address space. Bitlines are not segmented either in DRAM or in PCM. We observe that \tech{}'s performance is still better. On average, \tech{}'s execution time is 15\% lower than Baseline and 8\% lower than Nimble.

%% file: sections/related_works.tex
To our knowledge, this is the \textit{first} work that 1) enables segmented bitline architecture for non-volatile memory and analyzes its performance and reliability impacts, 2) develop a strategy to intelligently place pages in near and far segments of different memory units during their initial allocation, reducing run-time migration overhead, and 3) introduces page migrations within tiers of the same memory, without moving data on the memory channels.

\subsection{Performance, Energy, and Endurance Optimizations}
Many prior works optimize performance, energy, and endurance of PCM~\cite{LeeMicro10,yoon2015efficient,lee2010phase,QureshiISCA09,QureshiISCA12,ArjomandISCA16,cho2009flip,SeongSecurityISCA2010,akram2018write,songISMMa,song2019enabling}. Song et al. propose exploiting partition-level parallelism in each PCM bank to improve performance of DRAM-PCM hybrid memory \cite{song2019enabling}. Song et al. propose data content aware PCM writes to reduce write latency in PCM, improving performance of DRAM-PCM hybrid memory \cite{songISMMa}.  Cho et al. propose Flip-N-Write to improve PCM performance by first reading the memory content and then programming only the bits that need to be altered~\cite{cho2009flip}. Qureshi et al. propose PreSET, an architectural technique that SETs the PCM cells of a memory location in the background before programming them during write~\cite{QureshiISCA12}. 
There are also techniques to consolidate multiple write operations,
saving energy and improving performance~\cite{xia2014dwc} . 
As \tech{} addressees performance and reliability bottlenecks by tackling them at their source, it can be combined with these and similar techniques.

\subsection{Writeback Optimization}
Several prior works propose line-level writeback, where for each evicted DRAM cache block, processor cache blocks that become dirty are tracked and selectively written back to PCM~\cite{QureshiISCA09,LeeMicro10,lee2010phase,LeeISCA2009,pourshirazi2018wall,pourshirazi2019writeback}.
Various works propose dynamic write consolidation where PCM writes to the same row are consolidated into one write operation~\cite{xia2014dwc,StuecheliISCA,wang2013wade,lee2010dram,SeshadriISCA14}. 
Other works propose write activity reduction in PCM using CPU registers~\cite{hu2013write,huang2011register}.
Yet some other works propose multi-stage write operations where a write request is served in several steps rather than in one-shot to improve performance\cite{yue2013accelerating,zhang2016mellow}.
Qureshi et al.  propose a morphable PCM system, which dynamically adapts between high-density and high-latency MLC PCM and low-density and low-latency single-level cell PCM~\cite{qureshi2010morphable,qureshi2010improving}.
Jiang et al. propose write truncation where a write operation is truncated to allow read operations, compensating for data loss using ECC~\cite{jiang2012fpb}.
\tech~is complementary to all these approaches.

\subsection{Page Allocation}
Many modern OSes are already aware of 
performance and reliability characteristics of different memory technologies~\cite{lameter2013numa}. There are two different approaches for page placement in hybrid memory. The first approach is to monitor the memory
access patterns to pages, and migrate access-intensive pages to the faster memory, e.g., \cite{yan2019nimble,wang2019supporting,kannan2018heteroos,chang2018memory,kokolis2019pageseer,jayasena2018page}.
We compare \tech{} with Nimble~\cite{yan2019nimble} and found it to be largely better. 
However, the disadvantages of this approach are that it incurs high performance and energy overhead, as well as increasing
bank occupancy due to the movement of data in memory. An alternative approach is to predict the memory access patterns of pages and place them in matching memory during their initial allocation~\cite{khouzani2015improving}. However, it requires accurate predictions of the memory read-write characteristics of pages to be allocated. We not only introduce a new prediction scheme based on first-touch instruction, but also a novel memory architecture that can be exploited using this allocation policy. Furthermore, we discuss methods to reduce the migration overhead inside the memory bank.


%% file: sections/conclusions.tex
We introduce \tech{}, a new mechanism that enables segmented bitline architecture in DRAM-NVM hybrid memory, introducing \emph{intra- and inter-memory} performance and reliability asymmetries and exploit  them using an efficient page management policy at the OS, improving both performance and reliability. 
Previous architectural solutions exist to tackle page migration between different heterogeneous units in hybrid memory. However, they lead to significant performance and energy overhead due to high bank and channel occupancy during data migration.
In this paper, we first introduce an architectural solution involving the use of isolation transistors in long bitlines to create tiers with different latency and reliability characteristics. 
Next, we expose the asymmetric performance and reliability properties of memory tiers, both within and across heterogeneous memory units of hybrid memory to the OS. The OS exploits these asymmetries  in placing every newly-referenced memory page to the best tier during its initial allocation by predicting its data access intensity. This minimizes run-time page migrations, which lead to performance improvements. Finally, we propose a simple approach to facilitate page migrations within tiers of the same memory, eliminating the need to move data over memory channels, thereby further improving performance.
We evaluate \tech{} with single-core and multi-programmed workloads from the SPEC CPU2017 Benchmark suites.
Our results show that \tech{} significantly improves performance and reliability of state-of-the-art hybrid memory systems as well as mainstream DRAM-based systems. Additionally, \tech{}'s standalone page allocation policy can also be applied to improve performance of computing systems, where the proposed segmented bitline architecture is too costly to incorporate.  

We \textbf{conclude} that \tech{} is a \emph{simple yet powerful} mechanism for hybrid memory systems.

%% file: ismm20bCameraReady.bbl
\begin{thebibliography}{89}
\providecommand{\natexlab}[1]{#1}
\providecommand{\url}[1]{#1}
\csname url@samestyle\endcsname
\providecommand{\newblock}{\relax}
\providecommand{\bibinfo}[2]{#2}
\providecommand{\BIBentrySTDinterwordspacing}{\spaceskip=0pt\relax}
\providecommand{\BIBentryALTinterwordstretchfactor}{4}
\providecommand{\BIBentryALTinterwordspacing}{\spaceskip=\fontdimen2\font plus
\BIBentryALTinterwordstretchfactor\fontdimen3\font minus
  \fontdimen4\font\relax}
\providecommand{\BIBforeignlanguage}[2]{{%
\expandafter\ifx\csname l@#1\endcsname\relax
\typeout{** WARNING: IEEEtranSN.bst: No hyphenation pattern has been}%
\typeout{** loaded for the language `#1'. Using the pattern for}%
\typeout{** the default language instead.}%
\else
\language=\csname l@#1\endcsname
\fi
#2}}
\providecommand{\BIBdecl}{\relax}
\BIBdecl

\bibitem[jed(2019)]{jedecnvdimm2017}
``{Non-Volatile Dual In-line Memory Module (NVDIMM) -- N/F/P Specification},''
  \emph{JEDEC Solid State Technology Association}, 2019.

\bibitem[Akram et~al.(2018)Akram, Sartor, McKinley, and
  Eeckhout]{akram2018write}
S.~Akram, J.~B. Sartor, K.~S. McKinley, and L.~Eeckhout, ``Write-rationing
  garbage collection for hybrid memories,'' in \emph{PLDI}, 2018.

\bibitem[Antognetti and Massobrio(1990)]{antognetti1990semiconductor}
P.~Antognetti and G.~Massobrio, \emph{{Semiconductor device modeling with
  SPICE}}.\hskip 1em plus 0.5em minus 0.4em\relax McGraw-Hill, Inc., 1990.

\bibitem[Apalkov et~al.(2013)Apalkov, Khvalkovskiy, Watts, Nikitin, Tang,
  Lottis, Moon, Luo, Chen, Ong, et~al.]{apalkov2013spin}
D.~Apalkov, A.~Khvalkovskiy, S.~Watts, V.~Nikitin, X.~Tang, D.~Lottis, K.~Moon,
  X.~Luo, E.~Chen, A.~Ong \emph{et~al.}, ``{Spin-transfer torque magnetic
  random access memory (STT-MRAM)},'' \emph{JETC}, 2013.

\bibitem[Arafa and Ramanujan(2018)]{nation2009computer}
M.~Arafa and R.~K. Ramanujan, ``Memory card with volatile and non volatile
  memory space having multiple usage model configurations,'' \emph{\emph{US
  Patent 10,095,618}}, 2018.

\bibitem[Arjomand et~al.(2016)Arjomand, Kandemir, Sivasubramaniam, and
  Das]{ArjomandISCA16}
M.~Arjomand, M.~T. Kandemir, A.~Sivasubramaniam, and C.~R. Das, ``Boosting
  access parallelism to {PCM}-based main memory,'' in \emph{ISCA}, 2016.

\bibitem[Balaji et~al.(2019)Balaji, Song, Das, Dutt, Krichmar, Kandasamy, and
  Catthoor]{balajical19}
A.~Balaji, S.~Song, A.~Das, N.~Dutt, J.~Krichmar, N.~Kandasamy, and
  F.~Catthoor, ``A framework to explore workload-specific performance and
  lifetime trade-offs in neuromorphic computing,'' \emph{CAL}, 2019.

\bibitem[Bhattacharyya(2019)]{bhattacharyya2019memory}
A.~Bhattacharyya, ``Memory arrays,'' \emph{\emph{US Patent 10,374,101}}, 2019.

\bibitem[Blagodurov et~al.(2019)Blagodurov, Loh, and
  Meswani]{blagodurov2019hot}
S.~Blagodurov, G.~H. Loh, and M.~R. Meswani, ``Hot page selection in
  multi-level memory hierarchies,'' \emph{\emph{US Patent 10,235,290}}, 2019.

\bibitem[Bloom(1970)]{bloom1970space}
B.~H. Bloom, ``Space/time trade-offs in hash coding with allowable errors,''
  \emph{CSUR}, 1970.

\bibitem[Bourzac(2017)]{bourzac2017has}
K.~Bourzac, ``{Has Intel created a universal memory technology?}'' \emph{IEEE
  Spectrum}, 2017.

\bibitem[Bucek et~al.(2018)Bucek, Lange, et~al.]{bucek2018spec}
J.~Bucek, K.-D. Lange \emph{et~al.}, ``{SPEC CPU2017: Next-Generation Compute
  Benchmark},'' in \emph{ICPE}, 2018.

\bibitem[Cai et~al.(2017)Cai, Ghose, Haratsch, Luo, and Mutlu]{cai2017error}
Y.~Cai, S.~Ghose, E.~F. Haratsch, Y.~Luo, and O.~Mutlu, ``Error
  characterization, mitigation, and recovery in flash-memory-based solid-state
  drives,'' \emph{Proceedings of the IEEE}, 2017.

\bibitem[Cao and McAndrew(2007)]{cao2007mosfet}
Y.~Cao and C.~McAndrew, ``{MOSFET} modeling for 45nm and beyond,'' in
  \emph{ICCAD}, 2007.

\bibitem[Chandrasekar et~al.(2012)Chandrasekar, Weis, Li, Akesson, Wehn, and
  Goossens]{chandrasekar2012drampower}
K.~Chandrasekar, C.~Weis, Y.~Li, B.~Akesson, N.~Wehn, and K.~Goossens,
  ``{DRAMPower: Open-source DRAM power \& energy estimation tool},''
  \emph{http://www. drampower. info}, 2012.

\bibitem[Chang et~al.(2018)Chang, Chang, Chen, and Kuo]{chang2018memory}
Y.-M. Chang, Y.-H. Chang, H.-C. Chen, and T.-W. Kuo, ``Memory system and memory
  management method thereof,'' \emph{\emph{US Patent 10,108,555}}, 2018.

\bibitem[Cho et~al.(2005)Cho, Cho, Oh, and Choi]{cho2005programming}
B.-H. Cho, W.-Y. Cho, H.-R. Oh, and B.-G. Choi, ``Programming method of
  controlling the amount of write current applied to phase change memory device
  and write driver circuit therefor,'' \emph{\emph{US Patent 6,885,602}}, 2005.

\bibitem[Cho and Lee(2009)]{cho2009flip}
S.~Cho and H.~Lee, ``{Flip-N-Write: a simple deterministic technique to improve
  PRAM write performance, energy and endurance},'' in \emph{MICRO}, 2009.

\bibitem[Das and Kumar(2012)]{das2012fault}
A.~Das and A.~Kumar, ``Fault-aware task re-mapping for throughput constrained
  multimedia applications on noc-based mpsocs,'' in \emph{RSP}, 2012.

\bibitem[Das et~al.(2013{\natexlab{a}})Das, Kumar, and
  Veeravalli]{das2013aging}
A.~Das, A.~Kumar, and B.~Veeravalli, ``Aging-aware hardware-software task
  partitioning for reliable reconfigurable multiprocessor systems,'' in
  \emph{CASES}, 2013.

\bibitem[Das et~al.(2013{\natexlab{b}})Das, Kumar, and
  Veeravalli]{das2013reliability}
A.~Das, A.~Kumar, and B.~Veeravalli, ``Reliability-driven task mapping for
  lifetime extension of networks-on-chip based multiprocessor systems,'' in
  \emph{DATE}, 2013.

\bibitem[Das et~al.(2014{\natexlab{b}})Das, Kumar, and
  Veeravalli]{das2014communication}
A.~Das, A.~Kumar, and B.~Veeravalli, ``Communication and migration energy aware
  task mapping for reliable multiprocessor systems,'' \emph{FGCS}, 2014.

\bibitem[Das et~al.(2014{\natexlab{c}})Das, Kumar, and
  Veeravalli]{das2014energy}
A.~Das, A.~Kumar, and B.~Veeravalli, ``Energy-aware task mapping and scheduling
  for reliable embedded computing systems,'' \emph{TECS}, 2014.

\bibitem[Das et~al.(2014{\natexlab{a}})Das, Kumar, Veeravalli, Bolchini, and
  Miele]{das2014combined}
A.~Das, A.~Kumar, B.~Veeravalli, C.~Bolchini, and A.~Miele, ``Combined {DVFS}
  and mapping exploration for lifetime and soft-error susceptibility
  improvement in {MPSoCs},'' in \emph{DATE}, 2014.

\bibitem[Das et~al.(2015)Das, Kumar, and Veeravalli]{das2015reliability}
A.~Das, A.~Kumar, and B.~Veeravalli, ``Reliability and energy-aware mapping and
  scheduling of multimedia applications on multiprocessor systems,''
  \emph{TPDS}, 2015.

\bibitem[Das et~al.(2018)Das, Hassan, and Mutlu]{das2018vrl}
A.~Das, H.~Hassan, and O.~Mutlu, ``{VRL-DRAM: Improving DRAM performance via
  variable refresh latency},'' in \emph{DAC}, 2018.

\bibitem[Doweck et~al.(2017)Doweck, Kao, Lu, Mandelblat, Rahatekar, Rappoport,
  Rotem, Yasin, and Yoaz]{doweck2017inside}
J.~Doweck, W.-F. Kao, A.~K.-y. Lu, J.~Mandelblat, A.~Rahatekar, L.~Rappoport,
  E.~Rotem, A.~Yasin, and A.~Yoaz, ``{Inside 6th-generation Intel core: New
  microarchitecture code-named skylake},'' \emph{IEEE Micro}, 2017.

\bibitem[Dray and Wei(2018)]{dray2018high}
C.~Dray and L.~Wei, ``High voltage tolerant word-line driver,'' \emph{\emph{US
  Patent 9,875,783}}, 2018.

\bibitem[Fan et~al.(2014)Fan, Andersen, Kaminsky, and
  Mitzenmacher]{fan2014cuckoo}
B.~Fan, D.~G. Andersen, M.~Kaminsky, and M.~D. Mitzenmacher, ``{Cuckoo filter:
  Practically better than bloom},'' in \emph{CONEXT}, 2014.

\bibitem[Goda et~al.(2019)Goda, Vali, Miccoli, and
  Kalavade]{goda2018programming}
A.~Goda, T.~Vali, C.~Miccoli, and P.~Kalavade, ``Programming memory devices,''
  \emph{\emph{US Patent 10,217,515}}, 2019.

\bibitem[Hu et~al.(2013)Hu, Xue, Zhuge, Tseng, and Sha]{hu2013write}
J.~Hu, C.~J. Xue, Q.~Zhuge, W.-C. Tseng, and E.~H.-M. Sha, ``Write activity
  reduction on non-volatile main memories for embedded chip multiprocessors,''
  \emph{TECS}, 2013.

\bibitem[Huang et~al.(2011)Huang, Liu, and Xue]{huang2011register}
Y.~Huang, T.~Liu, and C.~J. Xue, ``Register allocation for write activity
  minimization on non-volatile main memory,'' in \emph{ASPDAC}, 2011.

\bibitem[Jayasena et~al.(2018)Jayasena, Loh, O'connor, and
  Chatterjee]{jayasena2018page}
N.~S. Jayasena, G.~H. Loh, J.~M. O'connor, and N.~Chatterjee, ``Page migration
  in a hybrid memory device,'' \emph{\emph{US Patent 9,910,605}}, 2018.

\bibitem[Jiang et~al.(2012)Jiang, Zhang, Childers, and Yang]{jiang2012fpb}
L.~Jiang, Y.~Zhang, B.~R. Childers, and J.~Yang, ``{FPB: Fine-grained power
  budgeting to improve write throughput of multi-level cell phase change
  memory},'' in \emph{MICRO}, 2012.

\bibitem[Kang et~al.(2014)Kang, Yu, Park, Zheng, Halbert, Bains, Jang, and
  Choi]{kang2014co}
U.~Kang, H.-s. Yu, C.~Park, H.~Zheng, J.~Halbert, K.~Bains, S.~Jang, and J.~S.
  Choi, ``Co-architecting controllers and {DRAM} to enhance {DRAM} process
  scaling,'' in \emph{The Memory Forum}, 2014.

\bibitem[Kannan et~al.(2018)Kannan, Gavrilovska, Gupta, and
  Schwan]{kannan2018heteroos}
S.~Kannan, A.~Gavrilovska, V.~Gupta, and K.~Schwan, ``{HeteroOS: OS} design for
  heterogeneous memory management in datacenters,'' \emph{OSR}, 2018.

\bibitem[Khouzani et~al.(2015)Khouzani, Yang, and Hu]{khouzani2015improving}
H.~A. Khouzani, C.~Yang, and J.~Hu, ``{Improving performance and lifetime of
  DRAM-PCM hybrid main memory through a proactive page allocation strategy},''
  in \emph{ASP-DAC}, 2015.

\bibitem[Kim et~al.(2012)Kim, Seshadri, Lee, Liu, and Mutlu]{KimISCA12}
Y.~Kim, V.~Seshadri, D.~Lee, J.~Liu, and O.~Mutlu, ``A case for exploiting
  subarray-level parallelism {(SALP) in DRAM},'' in \emph{ISCA}, 2012.

\bibitem[Kim et~al.(2016)Kim, Yang, and Mutlu]{kim2016ramulator}
Y.~Kim, W.~Yang, and O.~Mutlu, ``{Ramulator: A Fast and Extensible DRAM
  Simulator},'' \emph{CAL}, 2016.

\bibitem[Kokolis et~al.(2019)Kokolis, Skarlatos, and
  Torrellas]{kokolis2019pageseer}
A.~Kokolis, D.~Skarlatos, and J.~Torrellas, ``{PageSeer: Using page walks to
  trigger page swaps in hybrid memory systems},'' in \emph{HPCA}, 2019.

\bibitem[K{\"u}lt{\"u}rsay et~al.(2013)K{\"u}lt{\"u}rsay, Kandemir,
  Sivasubramaniam, and Mutlu]{kultursay2013evaluating}
E.~K{\"u}lt{\"u}rsay, M.~Kandemir, A.~Sivasubramaniam, and O.~Mutlu,
  ``Evaluating {STT-RAM} as an energy-efficient main memory alternative,'' in
  \emph{ISPASS}, 2013.

\bibitem[Lameter(2013)]{lameter2013numa}
C.~Lameter, ``Numa (non-uniform memory access): An overview,'' \emph{Queue},
  2013.

\bibitem[Lee et~al.(2010{\natexlab{b}})Lee, Zhou, Yang, Zhang, Zhao, Ipek,
  Mutlu, and Burger]{LeeMicro10}
B.~C. Lee, P.~Zhou, J.~Yang, Y.~Zhang, B.~Zhao, E.~Ipek, O.~Mutlu, and
  D.~Burger, ``Phase-change technology and the future of main memory,''
  \emph{IEEE Micro}, 2010.

\bibitem[Lee et~al.(2009{\natexlab{a}})Lee, Ipek, Mutlu, and Burger]{LeeISCA09}
B.~C. Lee, E.~Ipek, O.~Mutlu, and D.~Burger, ``Architecting phase change memory
  as a scalable dram alternative,'' in \emph{ISCA}, 2009.

\bibitem[Lee et~al.(2009{\natexlab{b}})Lee, Ipek, Mutlu, and
  Burger]{LeeISCA2009}
B.~C. Lee, E.~Ipek, O.~Mutlu, and D.~Burger, ``Architecting phase change memory
  as a scalable {DRAM} alternative,'' in \emph{ISCA}, 2009.

\bibitem[Lee et~al.(2010{\natexlab{c}})Lee, Ipek, Mutlu, and
  Burger]{lee2010phase}
B.~C. Lee, E.~Ipek, O.~Mutlu, and D.~Burger, ``Phase change memory architecture
  and the quest for scalability,'' \emph{CACM}, 2010.

\bibitem[Lee et~al.(2010{\natexlab{a}})Lee, Zhou, Yang, Zhang, Zhao, Ipek,
  Mutlu, and Burger]{lee2010phase2}
B.~C. Lee, P.~Zhou, J.~Yang, Y.~Zhang, B.~Zhao, E.~Ipek, O.~Mutlu, and
  D.~Burger, ``Phase-change technology and the future of main memory,''
  \emph{IEEE Micro}, 2010.

\bibitem[Lee et~al.(2010{\natexlab{d}})Lee, Narasiman, Ebrahimi, Mutlu, and
  Patt]{lee2010dram}
C.~J. Lee, V.~Narasiman, E.~Ebrahimi, O.~Mutlu, and Y.~N. Patt, ``{DRAM-aware
  last-level cache writeback: Reducing write-caused interference in memory
  systems},'' 2010.

\bibitem[Lee et~al.(2013)Lee, Kim, Seshadri, Liu, Subramanian, and
  Mutlu]{lee2013tiered}
D.~Lee, Y.~Kim, V.~Seshadri, J.~Liu, L.~Subramanian, and O.~Mutlu,
  ``{Tiered-latency DRAM: A low latency and low cost DRAM architecture},'' in
  \emph{HPCA}, 2013.

\bibitem[Luk et~al.(2005)Luk, Cohn, Muth, Patil, Klauser, Lowney, Wallace,
  Reddi, and Hazelwood]{LukPin}
C.-K. Luk, R.~Cohn, R.~Muth, H.~Patil, A.~Klauser, G.~Lowney, S.~Wallace, V.~J.
  Reddi, and K.~Hazelwood, ``Pin: Building customized program analysis tools
  with dynamic instrumentation,'' in \emph{PLDI}, 2005.

\bibitem[Lung et~al.(2016)Lung, Miller, Chen, Lewis, Morrish, Perri, Jordan,
  Ho, Chen, Chien, et~al.]{lung2016double}
H.-L. Lung, C.~P. Miller, C.-J. Chen, S.~C. Lewis, J.~Morrish, T.~Perri, R.~C.
  Jordan, H.-Y. Ho, T.-S. Chen, W.-C. Chien \emph{et~al.}, ``{A
  double-data-rate 2 (DDR2) interface phase-change memory with 533MB/s
  read-write data rate and 37.5 ns access latency for memory-type storage class
  memory applications},'' in \emph{IMW}, 2016.

\bibitem[Mallik et~al.(2017)Mallik, Garbin, Fantini, Rodopoulos, Degraeve,
  Stuijt, Das, Schaafsma, Debacker, Donadio, et~al.]{mallik2017design}
A.~Mallik, D.~Garbin, A.~Fantini, D.~Rodopoulos, R.~Degraeve, J.~Stuijt,
  A.~Das, S.~Schaafsma, P.~Debacker, G.~Donadio \emph{et~al.},
  ``Design-technology co-optimization for {OxRRAM}-based synaptic processing
  unit,'' in \emph{VLSI Technology}, 2017.

\bibitem[Mutlu(2013)]{mutlu2013memory}
O.~Mutlu, ``Memory scaling: A systems architecture perspective,'' in
  \emph{IMW}, 2013.

\bibitem[Mutlu(2017)]{mutlu2017rowhammer}
O.~Mutlu, ``The {RowHammer} problem and other issues we may face as memory
  becomes denser,'' in \emph{DATE}, 2017.

\bibitem[Mutlu and Kim(2019)]{mutlu2019rowhammer}
O.~Mutlu and J.~S. Kim, ``Rowhammer: A retrospective,'' \emph{TCAD}, 2019.

\bibitem[Mutlu and Subramanian(2015)]{mutlu2015research}
O.~Mutlu and L.~Subramanian, ``Research problems and opportunities in memory
  systems,'' \emph{Supercomputing Frontiers and Innovations}, 2015.

\bibitem[Nirschl et~al.(2007)Nirschl, Philipp, Happ, Burr, Rajendran, Lee,
  Schrott, Yang, Breitwisch, Chen, et~al.]{nirschl2007write}
T.~Nirschl, J.~Philipp, T.~Happ, G.~W. Burr, B.~Rajendran, M.-H. Lee,
  A.~Schrott, M.~Yang, M.~Breitwisch, C.-F. Chen \emph{et~al.}, ``Write
  strategies for 2 and 4-bit multi-level phase-change memory,'' in \emph{IEDM},
  2007.

\bibitem[Poremba et~al.(2015)Poremba, Zhang, and Xie]{poremba2015nvmain}
M.~Poremba, T.~Zhang, and Y.~Xie, ``Nvmain 2.0: A user-friendly memory
  simulator to model (non-) volatile memory systems,'' \emph{CAL}, 2015.

\bibitem[Pourshirazi et~al.(2018)Pourshirazi, Beigi, Zhu, and
  Memik]{pourshirazi2018wall}
B.~Pourshirazi, M.~V. Beigi, Z.~Zhu, and G.~Memik, ``{WALL: A writeback-aware
  LLC management for PCM-based main memory systems},'' in \emph{DATE}, 2018.

\bibitem[Pourshirazi et~al.(2019)Pourshirazi, Beigi, Zhu, and
  Memik]{pourshirazi2019writeback}
B.~Pourshirazi, M.~V. Beigi, Z.~Zhu, and G.~Memik, ``Writeback-aware {LLC}
  management for {PCM}-based main memory systems,'' \emph{TODAES}, 2019.

\bibitem[Qureshi et~al.(2009)Qureshi, Srinivasan, and Rivers]{QureshiISCA09}
M.~K. Qureshi, V.~Srinivasan, and J.~A. Rivers, ``Scalable high performance
  main memory system using phase-change memory technology,'' in \emph{ISCA},
  2009.

\bibitem[Qureshi et~al.(2010{\natexlab{a}})Qureshi, Franceschini,
  Lastras-Monta\~{n}o, and Karidis]{qureshi2010morphable}
M.~K. Qureshi, M.~M. Franceschini, L.~A. Lastras-Monta\~{n}o, and J.~P.
  Karidis, ``Morphable memory system: A robust architecture for exploiting
  multi-level phase change memories,'' in \emph{ISCA}, 2010.

\bibitem[Qureshi et~al.(2010{\natexlab{b}})Qureshi, Franceschini, and
  Lastras-Montano]{qureshi2010improving}
M.~K. Qureshi, M.~M. Franceschini, and L.~A. Lastras-Montano, ``Improving read
  performance of phase change memories via write cancellation and write
  pausing,'' in \emph{HPCA}, 2010.

\bibitem[Qureshi et~al.(2012)Qureshi, Franceschini, Jagmohan, and
  Lastras]{QureshiISCA12}
M.~K. Qureshi, M.~M. Franceschini, A.~Jagmohan, and L.~A. Lastras, ``{PreSET:}
  improving performance of phase change memories by exploiting asymmetry in
  write times,'' in \emph{ISCA}, 2012.

\bibitem[Redaelli(2018)]{pcm_book}
A.~Redaelli, ``{Phase Change Memory: Device Physics, Reliability and
  Applications},'' \emph{Phase Change Memory}, 2018.

\bibitem[Redaelli and Perrone(2017)]{redaelli2017semiconductor}
A.~Redaelli and C.~Perrone, ``Semiconductor constructions and memory arrays,''
  \emph{\emph{US Patent 9,748,480}}, 2017.

\bibitem[Ren et~al.(2015)Ren, Zhao, Khan, Choi, Wu, and Mutlu]{ren2015thynvm}
J.~Ren, J.~Zhao, S.~Khan, J.~Choi, Y.~Wu, and O.~Mutlu, ``{ThyNVM: Enabling
  software-transparent crash consistency in persistent memory systems},'' in
  \emph{MICRO}, 2015.

\bibitem[Rixner et~al.(2000)Rixner, Dally, Kapasi, Mattson, and
  Owens]{RixnerISCA2000}
S.~Rixner, W.~J. Dally, U.~J. Kapasi, P.~Mattson, and J.~D. Owens, ``Memory
  access scheduling,'' in \emph{ISCA}, 2000.

\bibitem[Sadasivam et~al.(2017)Sadasivam, Thompto, Kalla, and
  Starke]{sadasivam2017ibm}
S.~K. Sadasivam, B.~W. Thompto, R.~Kalla, and W.~J. Starke, ``{IBM Power9
  processor architecture},'' \emph{IEEE Micro}, 2017.

\bibitem[Sandhu et~al.(2020)Sandhu, Pietrzyk, and Lattimore]{sandhu2018memory}
B.~S. Sandhu, C.~Pietrzyk, and G.~M. Lattimore, ``Memory write driver, method
  and system,'' \emph{\emph{US Patent 10,529,420}}, 2020.

\bibitem[Seong et~al.(2010)Seong, Woo, and Lee]{SeongSecurityISCA2010}
N.~H. Seong, D.~H. Woo, and H.-H.~S. Lee, ``{Security Refresh: Prevent}
  malicious wear-out and increase durability for phase-change memory with
  dynamically randomized optaddress mapping,'' in \emph{ISCA}, 2010.

\bibitem[Seshadri et~al.(2013)Seshadri, Kim, Fallin, Lee, Ausavarungnirun,
  Pekhimenko, Luo, Mutlu, Gibbons, Kozuch, et~al.]{seshadri2013rowclone}
V.~Seshadri, Y.~Kim, C.~Fallin, D.~Lee, R.~Ausavarungnirun, G.~Pekhimenko,
  Y.~Luo, O.~Mutlu, P.~B. Gibbons, M.~A. Kozuch \emph{et~al.}, ``{RowClone:
  fast and energy-efficient in-DRAM bulk data copy and initialization},'' in
  \emph{MICRO}, 2013.

\bibitem[Seshadri et~al.(2014)Seshadri, Bhowmick, Mutlu, Gibbons, Kozuch, and
  Mowry]{SeshadriISCA14}
V.~Seshadri, A.~Bhowmick, O.~Mutlu, P.~B. Gibbons, M.~A. Kozuch, and T.~C.
  Mowry, ``The dirty-block index,'' in \emph{ISCA}, 2014.

\bibitem[Song et~al.(2019)Song, Das, Mutlu, and Kandasamy]{song2019enabling}
S.~Song, A.~Das, O.~Mutlu, and N.~Kandasamy, ``Enabling and exploiting
  partition-level parallelism ({PALP}) in phase change memories,'' \emph{TECS},
  2019.

\bibitem[Song et~al.(2020)Song, Das, Mutlu, and Kandasamy]{songISMMa}
S.~Song, A.~Das, O.~Mutlu, and N.~Kandasamy, ``Improving phase change memory
  performance with data content aware access,'' in \emph{ISMM}, 2020.

\bibitem[Srinivasan et~al.(2004)Srinivasan, Adve, Bose, and
  Rivers]{SrinivasanISCA04}
J.~Srinivasan, S.~V. Adve, P.~Bose, and J.~A. Rivers, ``The case for lifetime
  reliability-aware microprocessors,'' in \emph{ISCA}, 2004.

\bibitem[Stine et~al.(2007)Stine, Castellanos, Wood, Henson, Love, Davis,
  Franzon, Bucher, Basavarajaiah, Oh, et~al.]{stine2007freepdk}
J.~E. Stine, I.~Castellanos, M.~Wood, J.~Henson, F.~Love, W.~R. Davis, P.~D.
  Franzon, M.~Bucher, S.~Basavarajaiah, J.~Oh \emph{et~al.}, ``{FreePDK: An
  open-source variation-aware design kit},'' in \emph{MSE}, 2007.

\bibitem[Stuecheli et~al.(2010)Stuecheli, Kaseridis, Daly, Hunter, and
  John]{StuecheliISCA}
J.~Stuecheli, D.~Kaseridis, D.~Daly, H.~C. Hunter, and L.~K. John, ``The
  virtual write queue: Coordinating dram and last-level cache policies,'' in
  \emph{ISCA}, 2010.

\bibitem[Villa(2018)]{villa2018pcm}
C.~Villa, ``{PCM} array architecture and management,'' in \emph{Phase Change
  Memory}, 2018.

\bibitem[Wang et~al.(2019)Wang, Liu, Liao, Chen, Jin, Zhang, Zheng, He, and
  Jiang]{wang2019supporting}
X.~Wang, H.~Liu, X.~Liao, J.~Chen, H.~Jin, Y.~Zhang, L.~Zheng, B.~He, and
  S.~Jiang, ``Supporting superpages and lightweight page migration in hybrid
  memory systems,'' \emph{TACO}, 2019.

\bibitem[Wang et~al.(2013)Wang, Shan, Cao, Gu, Xu, Mu, Xie, and
  Jim{\'e}nez]{wang2013wade}
Z.~Wang, S.~Shan, T.~Cao, J.~Gu, Y.~Xu, S.~Mu, Y.~Xie, and D.~A. Jim{\'e}nez,
  ``{WADE: Writeback-aware dynamic cache management for NVM-based main memory
  system},'' \emph{TACO}, 2013.

\bibitem[Wilkes(2001)]{wilkes2001memory}
M.~V. Wilkes, ``The memory gap and the future of high performance memories,''
  \emph{Computer Architecture News}, 2001.

\bibitem[Wong et~al.(2010)Wong, Raoux, Kim, Liang, Reifenberg, Rajendran,
  Asheghi, and Goodson]{wong2010phase}
H.-S.~P. Wong, S.~Raoux, S.~Kim, J.~Liang, J.~P. Reifenberg, B.~Rajendran,
  M.~Asheghi, and K.~E. Goodson, ``Phase change memory,'' \emph{Proceedings of
  the IEEE}, 2010.

\bibitem[Wulf and McKee(1995)]{wulf1995hitting}
W.~A. Wulf and S.~A. McKee, ``Hitting the memory wall: implications of the
  obvious,'' \emph{Computer Architecture News}, 1995.

\bibitem[Xia et~al.(2014)Xia, Jiang, Xiong, Chen, Zhang, and Sun]{xia2014dwc}
F.~Xia, D.~Jiang, J.~Xiong, M.~Chen, L.~Zhang, and N.~Sun, ``{DWC: Dynamic
  write consolidation for phase change memory systems},'' in \emph{ICS}, 2014.

\bibitem[Yan et~al.(2019)Yan, Lustig, Nellans, and
  Bhattacharjee]{yan2019nimble}
Z.~Yan, D.~Lustig, D.~Nellans, and A.~Bhattacharjee, ``Nimble page management
  for tiered memory systems,'' in \emph{ASPLOS}, 2019.

\bibitem[Yoon et~al.(2015)Yoon, Meza, Muralimanohar, Jouppi, and
  Mutlu]{yoon2015efficient}
H.~Yoon, J.~Meza, N.~Muralimanohar, N.~P. Jouppi, and O.~Mutlu, ``Efficient
  data mapping and buffering techniques for multilevel cell phase-change
  memories,'' \emph{TACO}, 2015.

\bibitem[Yue and Zhu(2013)]{yue2013accelerating}
J.~Yue and Y.~Zhu, ``Accelerating write by exploiting pcm asymmetries,'' in
  \emph{HPCA}, 2013.

\bibitem[Zhang et~al.(2016)Zhang, Neely, Franklin, Strukov, Xie, and
  Chong]{zhang2016mellow}
L.~Zhang, B.~Neely, D.~Franklin, D.~Strukov, Y.~Xie, and F.~T. Chong, ``{Mellow
  writes: Extending lifetime in resistive memories through selective slow write
  backs},'' in \emph{ISCA}, 2016.

\end{thebibliography}
